\documentclass[a4paper,11pt]{article}
\pdfoutput=1
\usepackage{fullpage}
\usepackage{graphicx}
\usepackage{float}
\usepackage{rotating}
\usepackage{hyperref}
\usepackage[font={bf,small}]{caption}
\usepackage{amsfonts}
\usepackage{amssymb,amsmath}

\usepackage{authblk}

\usepackage{wrapfig}
\usepackage{enumerate}
\usepackage{titlesec}
\titleformat*{\section}{\Large\bfseries}
\titleformat*{\subsection}{\large\bfseries}
\titleformat*{\subsubsection}{\normalsize\bfseries}
\titlespacing*{\section} {0pt}{2ex plus 1ex minus .2ex}{1.5ex plus .2ex}
\titlespacing*{\subsection} {0pt}{1.5ex plus 1ex minus .2ex}{1ex plus .2ex}
\titlespacing*{\subsubsection}{0pt}{1.5ex plus 1ex minus .2ex}{1ex plus .2ex}
\usepackage{geometry}
\usepackage{enumitem}
\usepackage{xcolor}
\usepackage{soul}
 \DeclareMathSizes{10}{10}{9}{8}
 \usepackage{textgreek}

\usepackage[firstinits=true,backend=bibtex,doi=false,isbn=false,url=false,sorting=none,maxcitenames=1,maxbibnames=4]{biblatex}

\AtBeginBibliography{\footnotesize}
\renewbibmacro{in:}{}

\begin{document}
\title{{\bf{\LARGE A stochastic network approach to clustering\\ and visualising single-cell genomic count data}}}
\date{}
\author[1*]{Thomas E. Bartlett}
\author[2]{Swati Chandna}
\author[3]{Sandipan Roy}
\affil[1]{Department of Statistical Science, University College London, U.K.}
\affil[2]{School of Computing and Mathematical Sciences, Birkbeck, University of London, U.K.}
\affil[3]{Department of Mathematical Sciences, University of Bath, U.K.}
\affil[*]{thomas.bartlett.10@ucl.ac.uk}
\maketitle

\vspace{-6ex}

\begin{abstract}
Important tasks in the study of genomic data include the identification of groups of similar cells (for example by clustering), and visualisation of data summaries (for example by dimensional reduction). In this paper, we develop a novel approach to these tasks in the context of single-cell genomic data. To do so, we propose to model the observed genomic data count matrix $\mathbf{X}\in\mathbb{Z}_{\geq0}^{p\times n}$, by representing these measurements as a bipartite network with multi-edges. Utilising this first-principles network model of the raw data, we cluster single cells in a suitably identified {\it d}-dimensional Laplacian Eigenspace (LE) via a Gaussian mixture model (GMM-LE), and employ UMAP to non-linearly project the LE to two dimensions for visualisation (UMAP-LE). This LE representation of the data-points estimates transformed latent positions (of genes and cells), under a latent position statistical model of nodes in a bipartite stochastic network. We demonstrate how transformations of these estimated latent positions can enable fine-grained clustering and visualisation of single-cell genomic data, by application to data from three recent genomics studies in different biological contexts. In each data application, clusters of cells independently learned by our proposed methodology are found to correspond to cells expressing specific marker genes that were independently defined by domain experts. In this validation setting, our proposed clustering methodology outperforms the industry-standard for these data. Furthermore, we validate components of the LE decomposition of the data by contrasting healthy cells from normal and at-risk groups in a machine-learning model, thereby identifying an LE cancer biomarker that significantly predicts long-term patient survival outcome in two independent validation cohorts with data from 1904 and 1091 individuals.
\end{abstract}

\vspace{-1ex}

\section{Introduction}

\begin{figure}[!h]
\vspace{1ex}
\flushleft
{\LARGE (a)}\\
\vspace{-3ex}
\centering
\includegraphics[width=0.85\textwidth]{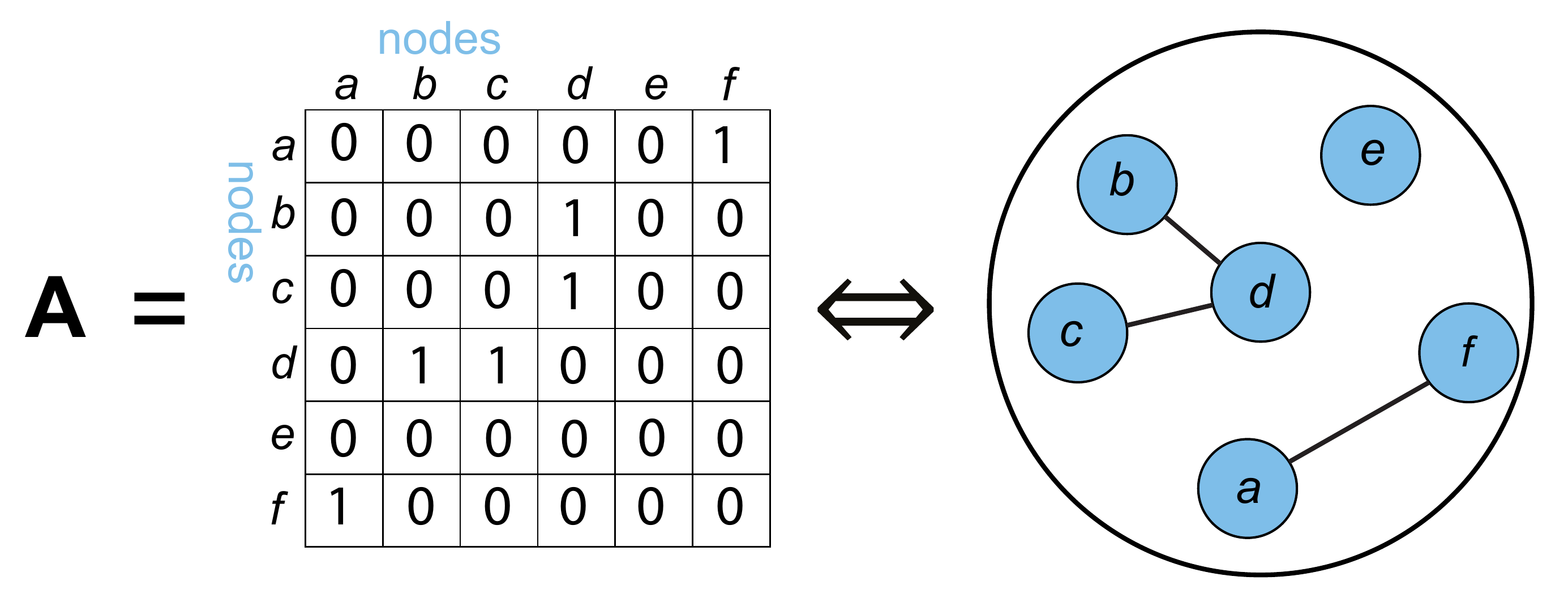}\\
\flushleft
\vspace*{8ex}
{\LARGE (b)}\\
\vspace*{-3ex}
\includegraphics[width=0.85\textwidth]{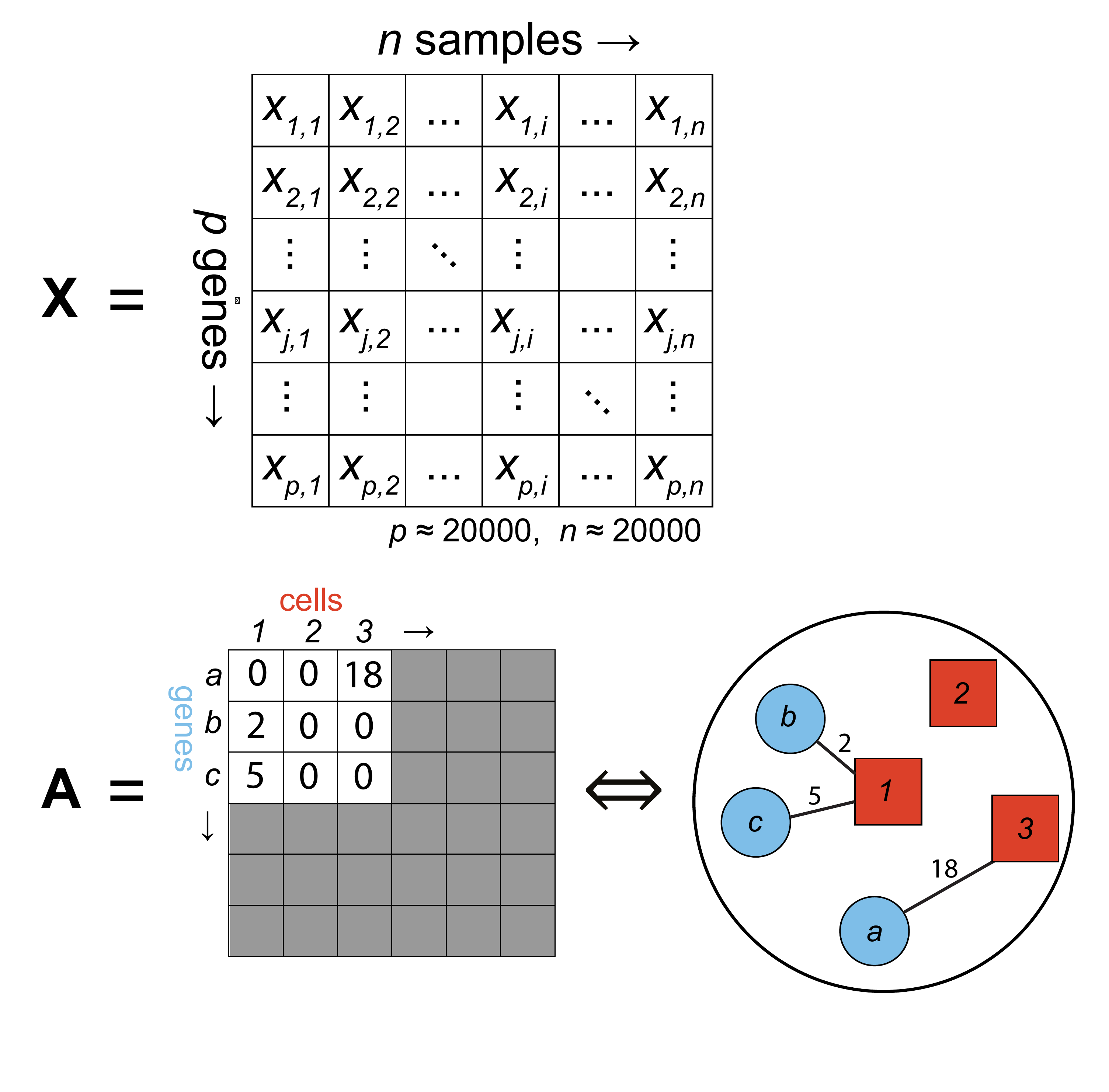}
\vspace*{-2ex}
\caption{Network model of count data. {\normalfont (a) Adjacency matrix $\mathbf{A}\in\{0,1\}^{n\times n}$ represents a symmetric unipartite network. (b) Adjacency matrix $\mathbf{A}\in\mathbb{Z}^{p\times n}_{\geq0}$ represents an asymmetric bipartite multi-edge network; adjacency matrix $\mathbf{A}$ (and the equivalent network it represents) models the data matrix $\mathbf{X}$ of non-negative integer counts.}}\label{introFig}
\vspace{1ex}
\end{figure}

Network models and methods have become an important subject in modern statistical science, especially in application-areas such as cell biology. A network can be represented by an `adjacency matrix', denoted by $\mathbf{A}$ (Figure \ref{introFig}). In the simplest case, a network consists of nodes of only one type (called a `unipartite network'), and these nodes may be connected (or not) by exactly one unweighted edge; in this case, adjacency matrix $\mathbf{A}\in\{0,1\}^{n\times n}$ (Figure \ref{introFig}a), where $n$ is the number of nodes in the network. When a network has nodes of two different types (represented by blue circles and red squares in Figure \ref{introFig}b), we call this a `bipartite' network; an example of a corresponding asymmetric adjacency matrix is $\mathbf{A}\in\{0,1\}^{p\times n}$. Network models are well established in genomics, and a stochastic network model \cite{erdHos1960evolution} refers to a statistical model of a network in which there is a probability $p_{ij}$ (defined according to some model) of observing an edge between nodes {\it i} and {\it j}. The edge-probability $p_{ij}$ is modelled according to a law which may depend on some data on or characteristics of the nodes {\it i} and {\it j} such as unobserved latent features $\phi_i^{(L)}$ and $\phi_j^{(R)}$ \cite{hoff2002latent,young2007random} (i.e., a latent position model), where we use the superscripts $\cdot^{(L)}$ and $\cdot^{(R)}$ to differentiate between vectors corresponding to the two different types of nodes in our bipartite setting. In this manuscript we focus on methodology that follows a latent position model, although we note that alternative stochastic network models exist, such as the exponential random graph model (ERGM) \cite{frank1986markov}.

In genomics, graphical models are often used to represent gene regulation in a gene regulatory network \cite{bray1995protein}, or related notions such as the gene co-expression network \cite{dobra2004sparse,kolar2010estimating}. Gene regulation is the phenomenon in which the protein encoded by one gene activates or represses the expression of another gene. When the expression of one gene regulates the expression of another gene in this way, this individual gene regulation is represented by an edge connecting the repressing and activating genes in the gene regulatory network. In contrast, the focus of the work presented in this paper is not graphical modelling of gene regulatory networks. It is important to make clear that in contrast to that literature, we use a mathematical object viz. a graph and its corresponding adjacency matrix directly as a mathematical-biological model \cite{murray2007mathematical} of single-cell genomic count data, in which numbers of mRNA molecules in specific cells are modelled by network edges. Then, we draw on recent advances in stochastic network theory in the mathematical-statistics literature to improve clustering and visualisation of the single cells in genomic data sets, based on our network representation of the data. Existing methods such as the industry-standard Seurat software package \cite{satija2015spatial} also model single-cell genomic count data with an adjacency matrix, by using the {\it k}-nearest-neighbours method to summarise the data matrix as a binary symmetric adjacency matrix $\mathbf{A}\in\{0,1\}^{n\times n}$. In contrast, we propose a first-principles representation of the data in which the raw data-counts of mRNA molecules in specific cells can be modelled directly with an asymmetric adjacency matrix $\mathbf{A}\in\mathbb{Z}_{\geq0}^{p\times n}$ with entries $A_{ij}$ that can take 0 or any positive integer. In other words, in our setup we propose modelling the data directly as a bipartite multi-edge network (Figure \ref{introFig}b).

Raw genomic sequencing data (RNA-seq, ATAC-seq, etc) may be represented as a matrix $\mathbf{X}\in\mathbb{Z}_{\geq0}^{p\times n}$ of non-negative integer counts. These are counts of genomic sequencing reads, such that each entry $X_{ij}$ represents the number of observed copies of genomic fragment (e.g., mRNA transcript) $i\in\{1,...,p\}$ in sample (e.g., cell) $j\in\{1,...,n\}$. Modern single-cell sequencing data use UMIs (unique molecular identifiers), which avoid double-counting the same transcript (after amplification by PCR, polymerase chain reaction). Historically, it was considered necessary to model dropouts in single-cell sequencing data, which manifest in the data as zero-inflation. However, it has been shown more recently that it is no longer necessary to model dropouts or zero-inflation if these data use UMIs \cite{townes2019feature}. For UMI data, under the model of \cite{townes2019feature} the columns of $\mathbf{X}$ are modelled from first-principles as $n$ independent draws from a multinomial distribution. This contrasts with long-established methods for analysing RNA-seq data which use normalisation methods such as TPM (transcripts per million) and FPKM (fragments per kilobase per million), as part of standardised pipelines. Normalisation methods such as TPM and FPKM normalise the data by library size (i.e., $\sum_{i=1}^pX_{ij}$), and this may induce spurious correlations between transcripts \cite{townes2019feature}, as well as making it difficult to write down expressions for the true error distributions of the data after normalisation. In other words, after normalisation, we can't say exactly what the true distribution of the normalised data is, and so rely on heuristics. There is currently no consensus in the genomics literature on best practice for normalising and modelling raw single-cell sequencing data \cite{hafemeister2019normalization,lause2021analytic,choudhary2022comparison}. Our proposed approach enables data modelling based on a first-principles representation of these data. This allows important insights into the distributions of representations of these data-points, such as asymptotic multivariate Gaussianity in spectrally-embedded space.

When the genomic data being analysed are the only available observations of some single cells in a study, an important task when analysing these data is to group cells together of similar types for identification or `phenotyping', e.g., by clustering. One of the most successful and reliable methods for clustering single cells in genomic data is the so-called `Louvain clustering' method \cite{blondel2008fast}, as implemented in the industry-standard Seurat software package \cite{satija2015spatial}. Seurat-Louvain clustering works by finding a {\it k}-nearest-neighbour graph and then optimising an estimator called the `network modularity', which is discussed in Section \ref{modSec}. Another important task when analysing single-cell data is visualisation of the data by projection into two dimensions, e.g., using non-linear methods such as UMAP (uniform manifold approximation and projection). Whilst care must be taken with interpretation of UMAP (and alternative non-linear) embeddings \cite{bohm2022attraction}, the main aim of these projections is that data points which are close together in the original high dimensional space should remain close together in the two dimensional projection. It follows from the foundational model of modern biology (the `central dogma of biology' \cite{crick1958protein}), that cells with similar gene expression profiles are considered to have similar phenotypes. So while directions may become distorted under projections such as UMAP, in general we can expect cells with similar phenotypes to remain located close together in the two dimensional projection. This means that UMAP projections of genomic data are very useful for visual validation and identification of clusters corresponding to specific cellular phenotypes. The {\it p} features of data matrix $\mathbf{X}\in\mathbb{R}^{p\times n}$ usually contain much redundancy, and so to aid computational tractability and reduce noise \cite{haber2017single}, spectral methods may be used to reduce the dimension of the data to $\mathbf{V}^\top\in\mathbb{R}^{d\times n}$, where $d\ll p$, before using a non-linear method such as UMAP to generate the two dimensional projection $\mathbf{Y}\in\mathbb{R}^{2\times n}$ from $\mathbf{V}^\top$ \cite{modell2021graph}. It is natural to carry out both spectral clustering and non-linear projection using the same spectral decomposition $\mathbf{V}$ of the data-matrix $\mathbf{X}$.

The structure of this papers is as follows. In Section \ref{methodsSect}, we present our novel first-principles model of single-cell genomic count data, and based on this our improved method of clustering single cells that we call `GMM-LE' clustering, and related visualisations that we call `UMAP-LE'. Then in Section \ref{resultsSect}, we present the results of applying GMM-LE clustering and UMAP-LE visualisation to example data-sets relevant to human cortical development, to human embryonic development, and to a population at risk of breast cancer. Finally, we identify and validate a machine-learning cancer biomarker obtained in a novel way from the LE representation of the data.

\section{Methods}\label{methodsSect}

\subsection{Model specification}
The mathematical-biological model specification that we propose here follows from the observation that a single-cell genomic data-matrix $\mathbf{X}\in\mathbb{Z}_{\geq0}^{p\times n}$ has equivalent characteristics to the adjacency matrix $\mathbf{A}$ that represents a bipartite network with multi-edges (Figure \ref{introFig}b). By `multi-edges' we mean that a pair of nodes may be connected by multiple edges, that is, $A_{ij}\in\mathbb{Z}_{\geq0}$. By `bipartite' we mean that the network can have nodes of two different types (row-nodes and column-nodes, represented by blue circles and red squares in Figure \ref{introFig}b), and in our setting we restrict this so that edges can only connect nodes of different types. Furthermore, both $\mathbf{A}$ and $\mathbf{X}$ tend to be very sparse (i.e., contain a large proportion of zeros: typically fewer than 10\% of the matrix elements are non-zero in both cases). Hence, we propose a mathematical framework where the numbers of mRNA molecules in specific cells are represented by a mathematical object viz. a graph and its corresponding adjacency matrix, i.e. $\mathbf{A}\in\mathbb{Z}_{\geq0}^{p\times n}$. Representing $\mathbf{X}$ by $\mathbf{A}$ hence provides a mathematical justification for using clustering methodology informed by stochastic network modelling, as follows.

\subsection{Model and inference details}
Having specified our representation of the genomic data-set as $\mathbf{A}=\mathbf{X}\in\mathbb{Z}_{\geq0}^{p\times n}$, we assume that the $A_{ij}$ in the corresponding bipartite network occur independently according to some probability mass function (PMF) $p_{A_{ij}}\in[0,1]$ as
\vspace*{-2ex}
\begin{equation}
\text{Pr}\left(A_{ij}\right)=p_{A_{ij}}\left(A_{ij};\phi^{(L)}_i,\phi^{(R)}_j\right);\enspace\sum_{A_{ij}=0}^\infty p_{A_{ij}}\left(A_{ij};\phi^{(L)}_i,\phi^{(R)}_j\right)=1,\enspace i\in\{1,...,p\},\enspace j\in\{1,...,n\},
\label{latentPosMod}
\vspace*{1ex}
\end{equation}
such that $E(A_{ij}|\phi^{(L)}_i,\phi^{(R)}_j)$ admits a low-rank representation. We note that in our setting, $A_{ij}$ is the number of edges between the $i^\text{th}$ row-node and the $j^\text{th}$ column-node in the network model, which corresponds to $X_{ij}$, the number of counts for the $i^\text{th}$ gene in the $j^\text{th}$ cell in the genomic data-set. Genomic count data are typically over-dispersed compared to a Poisson model \cite{townes2019feature,hafemeister2019normalization,lause2021analytic,choudhary2022comparison}, and hence we work under the more general setting of Eq.\ref{latentPosMod}.

A canonical example of the latent position model of Hoff \cite{hoff2002latent} with $A_{ij}\in\{0,1\}$ is $A_{ij}\sim\text{Bernoulli}(p_{ij})$ (which we note may be approximated by $A_{ij}\sim\text{Poisson}(p_{ij})$ \cite{perry2012null}). In that setting, our model would reduce to $E(A_{ij}|\phi_i,\phi_j)=p_{ij}=\text{Pr}\left(A_{ij}=1|\phi_i,\phi_j\right)$, which admits a low rank representation as often described in the literature under the random dot-product graph (RDPG) \cite{young2007random} and generalised random dot-product graph (GRDPG) \cite{rubin2022statistical} models, where $p_{ij}$ takes the form of an inner product and generalised inner product of latent position vectors, respectively. Under the (G)RDPG models, the latent positions can be estimated by the eigenvalues of the adjacency matrix scaled by the square-roots of the absolute values of their corresponding eigenvalues. More generally, under these models the latent position vectors $\phi^{(L)}_i$ and $\phi^{(R)}_j$ can be estimated consistently (up to an orthogonal transformation) \cite{sussman2012consistent,lyzinski2014perfect,lyzinski2016community,rubin2022statistical} by spectral clustering of the adjacency matrix (known as adjacency spectral embedding, ASE) or of the Laplacian matrix, known as Laplacian spectral embedding (LSE). In this context, a Gaussian mixture model is appropriate to carry out the spectral clustering of the nodes, because their latent positions in the ASE, or the transformations of these latent positions in the LSE, are known to follow multivariate Gaussian distributions \cite{rubin2022statistical}. In the LSE, the latent positions of the nodes are transformed as a result of the normalisation by the degree distributions of the nodes (or functions thereof) that takes place in the calculation of the graph Laplacian. Furthermore, while scaling by the square-root of the eigenvalue magnitudes is required to obtain the latent positions under the RDPG and GRDPG, such scaling will not change the asymptotic Gaussian assumption, because the effect of this scaling is to stretch the axes of the multivariate Gaussian distribution. It is also known that this asymptotic Gaussian assumption extends to weighted graphs \cite{gallagher2023spectral}, such as the multi-edge graphs that are the focus of this article. We therefore assume that the transformed latent positions of the nodes are asymptotically multivariate-Gaussian distributed in the LSE of the multi-edge graphs that we discuss here, as a subset of weighted graphs (i.e., with weights taking only integer values). Hence for the unsupervised learning and visualisation applications described in this article, we do not need to specify the precise form of the PMF of Eq.\ref{latentPosMod}, such as the overdispersion of the edge-counts $A_{ij}$.

For genomic data, we choose to work with the Laplacian matrix as it is known to allow discovery of `affinity' structure (in the embedded space), in contrast with the ASE which is preferred for discovery of `core-periphery' structure \cite{priebe2019two}. In our proposed modelling setup for genomic data, affinity structure corresponds to grouping genes with similar patterns of expression across cells (via transformations of $\phi^{(L)}_i$ as the rows of $\mathbf{U}$), or to grouping cells with similar patterns of expression across genes (via transformations of $\phi^{(R)}_j$ as the rows of $\mathbf{V}$), i.e., cells with similar phenotypes. Because different dimensions of the spectral decomposition typically represent different aspects of cell-type identity (compare Fig.\ref{breastCancerDataFig} and Fig. \ref{pcScoresFig}), in the practical application of this methodology for unsupervised learning of cell-types we do not scale by the square-root of the eigenvalues, to enable discovery of novel cell-types which may be present in small numbers. We refer to the eigenspace of the Laplacian matrix as the `Laplacian Eigenspace' (LE), which is used to estimate transformations of latent position vectors $\phi^{(L)}_i$ and $\phi^{(R)}_j$ via singular value decomposition (SVD) of the graph-Laplacian $\boldsymbol{\mathcal{L}}\in\mathbb{R}_{\geq0}^{p\times n}$ as follows. We calculate the normalised graph-Laplacian as $\boldsymbol{\mathcal{L}}=\mathbf{D}_L^{-1/2}\mathbf{A}\mathbf{D}_R^{-1/2}$, where $\mathbf{D}_L$ and $\mathbf{D}_R$ are the diagonal matrices of degree distributions $\mathbf{d}_L$ and $\mathbf{d}_R$ in which the $i^\text{th}$ and $j^\text{th}$ elements are defined as $\sum_{j=1}^nA_{ij}$ and $\sum_{i=1}^pA_{ij}$ respectively \cite{rohe2016co,bartlett2021co}. Then we find the SVD as $\boldsymbol{\mathcal{L}}=\mathbf{U}\mathbf{S}\mathbf{V}^\top$, where $\mathbf{S}$ is a $d\times d$ diagonal matrix of singular values (SVs) with $S_{i,i}>0$, $i\in\{1,...d\}$, $\mathbf{U}\in\mathbb{R}^{p\times d}$ and $\mathbf{V}\in\mathbb{R}^{n\times d}$ are orthogonal matrices with their columns made up of left and right singular vectors (LSVs and RSVs) respectively, and $d$ is the dimensionality of the estimated spaces of both $\phi^{(L)}_i$ and $\phi^{(R)}_j$. Hence, we estimate transformations of $\phi^{(L)}_i$ and $\phi^{(R)}_j$ as {\it d}--dimensional vectors, directly as the rows of $\mathbf{U}$ and $\mathbf{V}$ respectively.

We assume that cells with similar patterns of gene expression represent cells with similar phenotypes. Therefore, clustering cells (samples) in the space of the RSVs (denoted $\mathbf{V}$) provides a sensitive method for unsupervised learning of known and unknown cell-types from unlabelled single-cell genomic data. In other words, we cluster together samples $j$ that have similar (transformed) latent positions $\phi^{(R)}_j$. We optimise this unsupervised learning procedure by following recent advances in understanding of the latent geometry of the LE that follow from the theory of stochastic networks. Following the central limit theorem established for (transformations of) latent position vectors estimated via LSE \cite{athreya2016limit,tang2018limit,rubin2022statistical}, we carry out this clustering by fitting a Gaussian mixture model in the LE and refer to it as GMM-LE clustering. Before clustering, we also normalise the rows of $\mathbf{U}$ and $\mathbf{V}$ to unit length (so that the transformed latent positions are projected onto a unit hypersphere), to help take account of nodes with very small degrees \cite{ng2001spectral,rohe2011spectral,lei2015consistency}. This normalisation is carried out after calculating `leverage scores' \cite{qin2013regularized} for each $j\in\{1,..n\}$ as $||\mathbf{v}_j||_2$, for use downstream in the GMM-LE clustering (Section \ref{sectSCapp}), where $\mathbf{v}_j$ is the $j^\text{th}$ row of $\mathbf{V}$. If necessary, we also inflate the degree distributions $\mathbf{d}_L$ and $\mathbf{d}_R$ by their medians to regularise the spectral clustering \cite{chaudhuri2012spectral,qin2013regularized,amini2013pseudo}, to improve computational identifiability in the presence of many low-degree nodes.

We also note that the random-walk Laplacian is relevant to our stochastic networks view of genomic network data, as follows. The random-walk Laplacian \cite{modell2021spectral} is calculated as $\boldsymbol{\mathcal{L}}=\mathbf{A}\mathbf{D}_R^{-1}$, which is equivalent to normalising the genomic data for each sample by its library size (a typical pre-processing step in genomics). This normalisation is also equivalent to discarding the top component of the spectral embedding (corresponding to the largest singular value of the SVD). Hence in this setting, the ASE with the top component discarded is equivalent to the spectral embedding of the random-walk Laplacian. We found that in practice, in addition to the ASE, the top component of the LSE of the genomic data count matrix is always highly correlated with the library sizes of that data set (\ref{topRSVcor}). Hence, we found that the normalised Laplacian with the top component discarded worked well with genomic data, in line with standard practice in genomics of normalising by library size, whilst also making the most of the favourable properties of the LE for finding features of interest in this genomics setting. Therefore, for subsequent clustering and visualisation steps, we calculate the normalised Laplacian and discard the top component.

\subsection{Related methods}\label{modSec}
We carry out spectral clustering of the raw genomic count matrix by fitting a Gaussian mixture model (GMM) in the LE, which we call GMM-LE clustering. We start with a latent position model (Eq.\ref{latentPosMod}) of our bipartite graph representation of the count-data matrix $\mathbf{X}$, i.e. $\mathbf{A}=\mathbf{X}\in\mathbb{Z}_{\geq0}^{p\times n}$, noting the asymmetric adjacency matrix $\mathbf{A}$. Hence, our proposed methodology allows separate but closely connected clustering and analysis (Section \ref{sectSCapp}) of the two different types of nodes (corresponding to latent positions $\phi^{(L)}_i$, $i\in\{1,...,p\}$ and $\phi^{(R)}_j$, $j\in\{1,...,n\}$). This separate clustering of nodes of two types is referred to as co-clustering, or co-community detection \cite{rohe2016co,bartlett2021co}. Our approach is related to other recent approaches such as \cite{modell2022spectral}, in which a bipartite graph refers to one in which nodes may be partitioned into two different intrinsic groups. Those groups correspond to the two different types of nodes in our setup, which can be identified equivalently by spectral embedding of a symmetric adjacency matrix that has all nodes of all types on both margins, i.e.
$
\left[ \begin{array}{cc}
\mathbf{0} & \mathbf{A}^\top  \\
\mathbf{A} & \mathbf{0} \end{array} \right],
$
where $\mathbf{A}\in\mathbb{Z}_{\geq0}^{p\times n}$ is our asymmetric adjacency matrix $\mathbf{A}=\mathbf{X}$, and $\mathbf{0}$ is an appropriately-sized matrix of 0s. Such a matrix always has eigenvalues in pairs of equal magnitude and opposite sign, and negative eigenvalues are known to indicate disassortative features \cite{khor2010concurrency}, which in our context allow separation of the two different types of nodes (genes and cells). For example, consider the 2-node bipartite network represented (in our setting) by the adjacency matrix $A=[1]$, i.e., the corresponding data matrix $X=[1]$ represents 1 count of 1 gene in 1 cell. Then, the symmetric embedding of $A$ described above would be 
$
\left[ \begin{array}{cc}
0 & 1  \\
1 & 0 \end{array} \right],
$
which has eigenvalues 1 and -1, with corresponding eigenvectors $[1/\sqrt{2},1/\sqrt{2}]^\top$ and $[-1/\sqrt{2},1/\sqrt{2}]^\top$. The second of these eigenvectors is disassortative, because dividing the nodes based on a separating point anywhere in $\left(-1/\sqrt{2},1/\sqrt{2}\right)$ will separate the nodes into two groups according to their types `cell' and `gene' as a `cut' based on this second eigenvector \cite{shi2000normalized,von2007tutorial}. Whereas there is no separating point for which the first eigenvector can be used to separate the nodes into groups.

\begin{figure}[!ht]
\vspace{-1ex}
\includegraphics[width=0.96\textwidth]{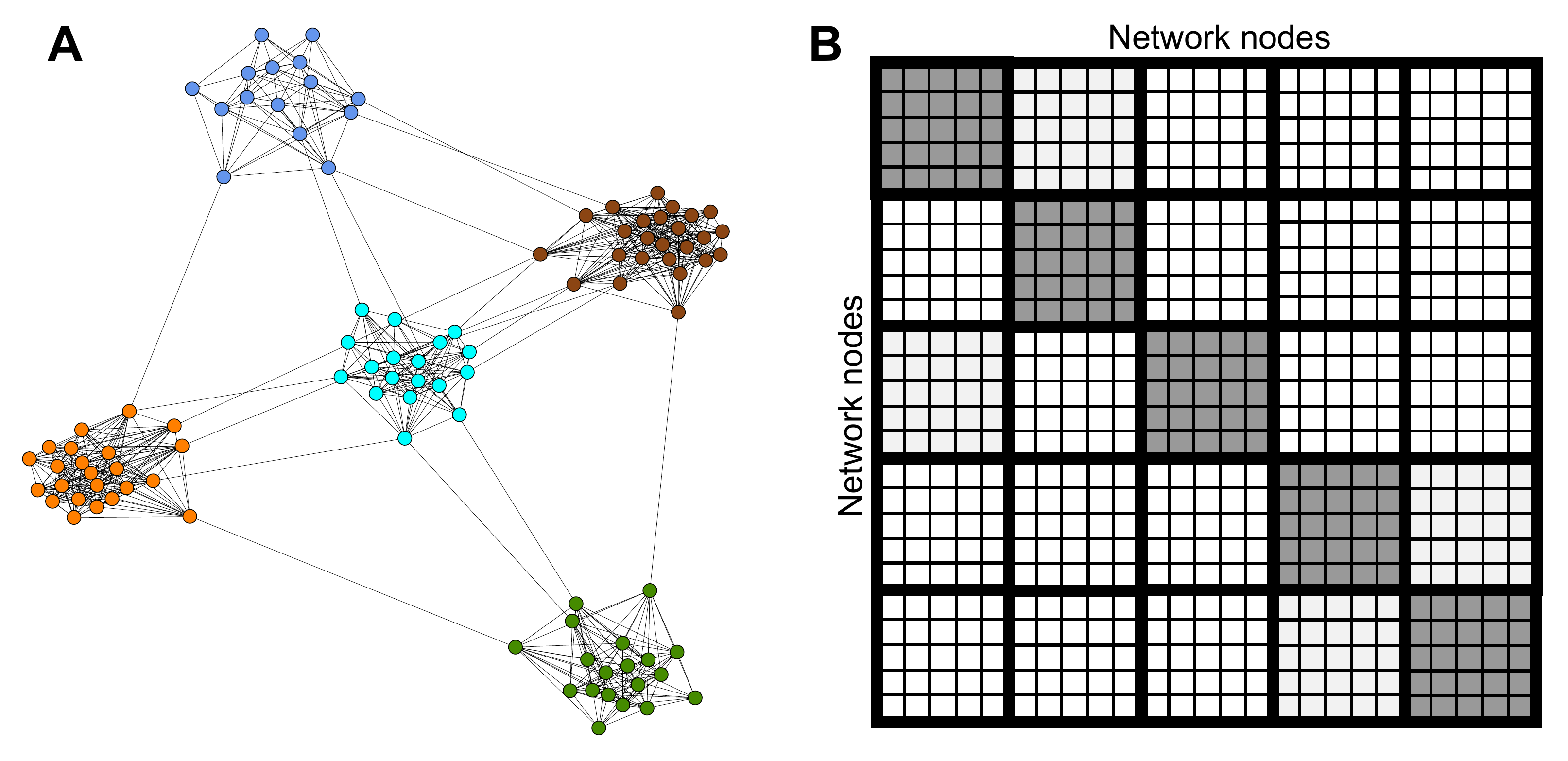}
\vspace{-1ex}
\caption{Communities in a symmetric adjacency matrix. {\normalfont (a) Network nodes in the same community are displayed with the same colour. The density of network edges tends to be much greater within communities, than between communities. (b) Network nodes are ordered by community (indicated by shading) along both margins, revealing the blockmodel structure.}}\label{communityFig}
\vspace{-2ex}
\end{figure}

Negative eigenvalues are not found when carrying out a singular value decomposition of an asymmetric adjacency matrix $\mathbf{A}\in\mathbb{Z}_{\geq0}^{p\times n}$ or corresponding asymmetric Laplacian $\mathcal{L}\in\mathbb{R}^{p\times n}$, because with $\boldsymbol{\mathcal{L}}=\mathbf{U}\mathbf{S}\mathbf{V}^\top$, it is necessary to enforce $S_{l,l}>0$, $l\in\{1,...d\}$, because the signs of $\mathbf{S}$ and $\mathbf{U}$ or of $\mathbf{S}$ and $\mathbf{V}$ could be arbitrarily switched. This also illustrates how in this asymmetric setting, different pairs of latent feature $\phi^{(L)}_i$, $i\in\{1,...,p\}$ with latent feature $\phi^{(R)}_j$, $j\in\{1,...,n\}$ may create assortative or disassortative effects, where we have {\em extended this terminology} for our setting, as follows. We identify effects that we term {\em assortative} or {\em disassortative} via the corresponding signs of the elements of $\mathbf{U}$ and $\mathbf{V}$, where pairs of groupings of nodes of each type can be represented as co-communities \cite{bartlett2021co}. \begin{wrapfigure}{r}{0.35\textwidth}
\vspace{-3ex}
\begin{center}
\includegraphics[width=0.35\textwidth]{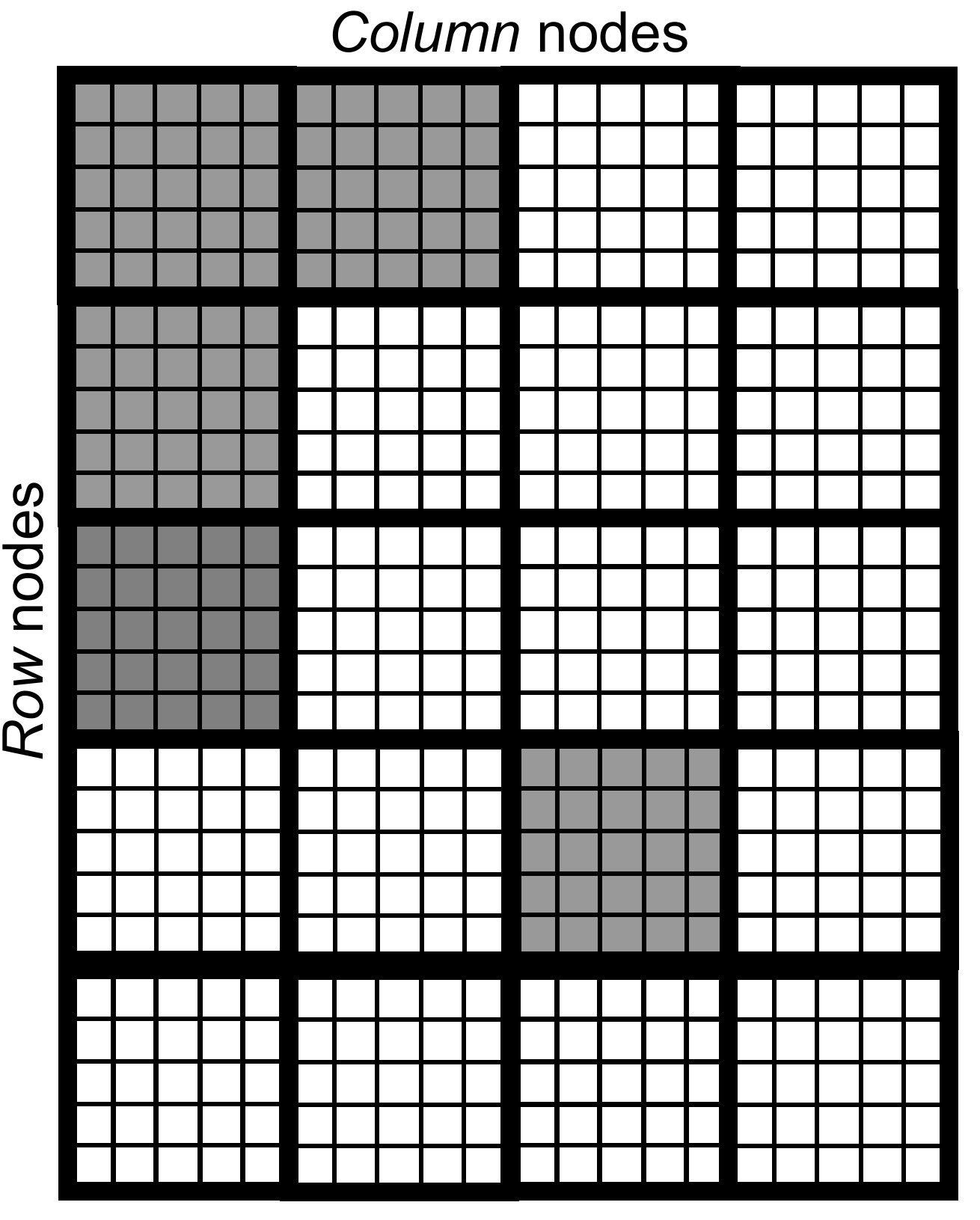}
\end{center}
\vspace{-2ex}
\caption{Co-communities in an asymmetric adjacency matrix. {\normalfont The margins of the adjacency matrix represent different types of nodes, and the adjacency matrix elements define the edges of the bipartite network. The shaded blocks represent co-communities of the two different types of nodes.}} \label{coComBaseEg}
\vspace{-5ex}
\end{wrapfigure} For example, consider the SVD decompositions $A_{ij}=\sum_{l=1}^dU_{i,l}S_{l,l}V_{l,j}$ and $A_{ij'}=\sum_{l=1}^dU_{i,l}S_{l,l}V_{l,j'}$ (for nodes $j$ and $j'$). We wish to cluster together cells (represented by transformations of the latent positions modelled by the rows of $\mathbf{V}$) if those cells have large counts of transcripts from similar sets of genes (i.e., similar gene expression profiles), indicating similar phenotypes. Then if $A_{ij}$ represents the number of genomic counts $X_{ij}$, consider the contribution to counts $X_{ij}$ and $X_{ij'}$ from SVD component $l$ in cases 1 and 2 below. Here we define gene $i$ as a marker gene for a certain cell type, $X_{ij}$ and $X_{ij'}$ as the counts for marker gene $i$ in cells $j$ and $j'$, and assume $U_{i,l}>0$:
\begin{enumerate}
\item $V_{l,j}>0$ and $V_{l,j'}>0$
\item $V_{l,j}>0$ and $V_{l,j'}<0$.
\end{enumerate}
\noindent
In case 1 above, feature $l$ makes a positive contribution to the count data for marker gene $i$ in both cells $j$ and $j'$; so we say that feature $l$ is {\it assortative} for cells $j$ and $j'$ (promoting their inclusion in the same cluster). But in case 2, now cells $j$ and $j'$ make positive and negative contributions (respectively) to the count data for marker gene $i$; so we say that feature $l$ is {\it disassortative} for cells $j$ and $j'$ (promoting their inclusion in different clusters).

The stochastic blockmodel \cite{holland1983stochastic,karrer2011stochastic,qin2013regularized} is another latent position model, under which community detection is carried out by spectral clustering \cite{newman2006modularity,qin2013regularized} and modularity maximisation \cite{newman2006modularity}. Communities in a network are groups of highly interconnected nodes in that network (Figure \ref{communityFig}a); ordering the nodes by community membership along the margins of the adjacency matrix reveals the blockmodel structure and modular nature of the network (Figure \ref{communityFig}b). Very importantly,  Newman equated spectral clustering of network nodes with maximising an estimator called the network modularity \cite{newman2006modularity}, unifying these different perspectives on equivalent methodologies. Network modularity is of particular relevance to genomic data, because modularity maximisation with a symmetric adjacency matrix $\mathbf{A}$ is the basis of the popular, reliable, and effective `Louvain clustering' algorithm \cite{blondel2008fast}, as implemented in the Seurat package which is widely used in genomics \cite{satija2015spatial}. This symmetric $\mathbf{A}\in\{0,1\}^{n\times n}$ is generated in Seurat-Louvain clustering from a {\it k}-nearest-neighbours graph derived from the data-matrix, in contrast to our first-principles model of the data-matrix $\mathbf{A}=\mathbf{X}\in\mathbb{Z}^{p\times n}_{\geq0}$. Other ways to generate a symmetric adjacency matrix from genomic data matrix $\mathbf{X}$ include thresholding the sample correlation matrix, based on prior knowledge or principled transformations \cite{bartlett2017network}.

In the bipartite setting, the comodularity is an analogous concept to the modularity \cite{bartlett2021co}. The comodularity has a fundamental difference in that it is calculated from $K^{(R)}\times K^{(C)}$ co-communities that may be located anywhere in an adjacency matrix (Figure \ref{coComBaseEg}, noting that the node locations have been ordered by row- and column-community membership). Whereas the modularity is calculated from $K$ communities that are located on the leading diagonal of the adjacency matrix (Figure \ref{communityFig}, noting that now the nodes have been ordered symmetrically by community membership). 

To summarise, in the results that we present in Section \ref{resultsSect} we focus on GMM-LE clustering for one of the node-types described (cells), i.e., clustering the rows of $\mathbf{V}\in\mathbb{R}^{n\times d}$. We visualise these nodes (cells) in $\mathbf{Y}^\top\in\mathbb{R}^{n\times 2}$ with UMAP-LE, where each sample-point $j\in\{1,...,n\}$ in the visualisation can be coloured as in Figures \ref{neuroDataFig}, \ref{embryoDataFig}, and \ref{breastCancerDataFig}. Nodes of only one type (i.e., cells) are plotted as dots in those figures. Each of these nodes/dots is then coloured based on the number of network edges that connect it to the other type of node (i.e., genes). Hence, the colour intensity of the dot plotted for each node (cell) $j$ is proportional to the number of connections $\sum_{i\in g} X_{ij}$ in the network $\mathbf{A}=\mathbf{X}\in\mathbb{Z}^{p\times n}_{\geq0}$ that node $j$ has to a predefined set of markers $i\in g$ of the other node-type (genes). We also use a machine-learning model to identify a cancer-predictive component $l\in\{1,...d\}$ of the $d$-dimensional LE via its assortative effect in the transformations of latent positions $\phi^{(R)}_j$, $j\in\{1,...,n\}$, as estimated by $V_{l,j}$. Then we validate a cancer biomarker based on $U_{l,i}$, where $i\in g'$ is from a set of genes $g'$ chosen by similar reasoning to the discussion of cases 1 and 2 above.

\subsection{Single-cell clustering}\label{sectSCapp}
Under the RDPG and GRDPG models, we expect asymptotically multivariate-Gaussian distributions of data-points in their transformed latent positions \cite{athreya2016limit,tang2018limit,rubin2022statistical}. Hence, we cluster cells using a Gaussian mixture model (GMM) in the Laplacian eigenspace (LE) $\mathbf{V}\in\mathbb{R}^{n\times K'}$ which we refer to as GMM-LE clustering, as follows. We note that a popular choice of $K'$ is $K'=K-1$ components, where $K$ is the number of clusters sought \cite{riolo2014first}.
\begin{enumerate}
\item For each cell/sample $j$, calculate their `leverage scores' \cite{qin2013regularized} as $||\mathbf{v}_j||_2$, where $\mathbf{v}_j$ is the $j^\text{th}$ row of $\mathbf{V}$, before normalising the rows of $\mathbf{V}$ to have unit length, i.e., calculate $\tilde{\mathbf{v}}_j=\mathbf{v}_j/||\mathbf{v}_j||_2$, $j\in\{1,...,n\}$.
\item Estimate model parameters $\hat{p}(k)$, $\hat{\boldsymbol{\mu}}_k$, and $\hat{\boldsymbol{\Sigma}}_k$, $k\in\{1,...,K\}$, by fitting a GMM (weighted by leverage score) with the expectation-maximisation (EM) algorithm to normalised data-points $\tilde{\mathbf{v}}_j$ for cells $\{j:||\mathbf{v}_j||_2\geq1/\sqrt{n}\}$.
\item Assign each cell-sample $j\in\{1,...,n\}$ to cluster $k\in\{1,...,K\}$ according to:\\ $k|j=\text{argmax}_k\hspace{0.5ex}\hat{p}(k|j)=\text{argmax}_k\left[\hat{p}(j|k)\hat{p}(k)\right]$, where $\hat{p}(j|k)=f_{\mathcal{N}}(\tilde{\mathbf{v}}_j|\hat{\boldsymbol{\mu}}_k,\hat{\boldsymbol{\Sigma}}_k)$,
\end{enumerate}
where $f_{\mathcal{N}}$ denotes Gaussian density. We note that clusters inferred using a Gaussian mixture model with diagonal covariance matrices (i.e., diagonal $\boldsymbol{\Sigma}_k$, $k\in\{1,...,K\}$) would be equivalent to the clusters obtained from the {\it K}-means algorithm.

\subsection{Single-cell data visualisaton}\label{sectSCvis}
Single-cell data are often visualised in two dimensions by using non-linear methods such as UMAP, i.e., projecting the data $\mathbf{X}\in\mathbb{Z}_{\geq0}^{p\times n}$ to give $\mathbf{Y}\in\mathbb{R}^{2\times n}$. To aid computational tractability and reduce noise, spectral decomposition is often applied to the data matrix to give $\mathbf{V}^\top\in\mathbb{R}^{d\times n}$, $d\ll p$, before carrying out non-linear projection to $\mathbf{Y}\in\mathbb{R}^{2\times n}$ for visualisation \cite{haber2017single}. We suggest that when clustering the data-points in their LE projection, to generate related visualisations then it is most natural to obtain a two-dimensional projection $\mathbf{Y}$ of these data-points by applying non-linear dimension reduction directly to the same LE projection, i.e., to the RSVs $\mathbf{V}$ of $\boldsymbol{\mathcal{L}}$. Representing single-cell genomic data as a bipartite network $\mathbf{A}=\mathbf{X}\in\mathbb{Z}_{\geq0}^{p\times n}$ also provides theoretical backing to the decision to estimate UMAP projections directly from LE projections, because recent work has shown that data-points projected into the LE will lie close to a low-dimensional manifold in a higher-dimensional ambient space \cite{rubin2020manifold,whiteley2021matrix}. In Section \ref{resultsSect}, we present visualisations of various single-cell genomic data-sets that follow from applying UMAP to the LE projection $\mathbf{V}$ of the data points, a procedure that we refer to as UMAP-LE (uniform manifold approximation and projection from the Laplacian eigenspace).

\section{Results}\label{resultsSect}
In this section, we use the methodology described in Section \ref{methodsSect}, to cluster and visualise single-cell genomic data from three recent studies from different cell-biological contexts:
\begin{itemize}
\item {\it Human cortical development} \cite{bhaduri2021atlas}: data for $p=15735$ RNA transcripts in $n=31073$ cells, from the V1 area of the visual cortex in the brains of human embryos at gestation weeks 20-22.
\item {\it Human embryonic development} \cite{petropoulos2016single}: data for $p=20407$ RNA transcripts in $n=1258$ cells, from the blastocyst stage of the embryo at 5-7 days post fertilisation.
\item {\it Breast cancer at-risk} \cite{pal2021single}: data for $p=14835$ RNA transcripts in $n=17730$ cells, from the epithelial lineage and supporting cells, from human breast tissue from healthy individuals.
\end{itemize}
Figure \ref{neuroDataFig}, Figure \ref{embryoDataFig}, and Figure \ref{breastCancerDataFig} show clustering and visualisation based on the Laplacian eigenspace (LE) transformed latent positions estimated for the data-points/cells of, respectively, the human cortical development data-set, the human embryonic development data-set, and the breast cancer at-risk data-set. UMAP is used to project the {\it n} data-points into two dimensions for visualisation, based on their transformed latent positions in the estimated Laplacian eigenspace (UMAP-LE). GMM-LE clustering is also carried out by fitting a Gaussian mixture model in the same estimated LE. In the plots in these figures, each dot or data-point represents one cell (sample). In the cluster plots, the colours indicate the GMM-LE cluster that the cell is assigned to. In the remaining plots, the colours indicate the average log-expression levels of sets of marker genes defined by domain experts for specific cell-types that are relevant to each data-set. These sets of marker genes have been used previously by us in other studies relating to human neural development \cite{bartlett2021two}, human embryonic development \cite{alanis2023mica}, and breast cancer at-risk \cite{bartlett2021inference}. 

\subsection{Human cortical development dataset: visualisation and clustering inferences}

\begin{figure}[!h]
\vspace{-1ex}
\includegraphics[width=0.96\textwidth]{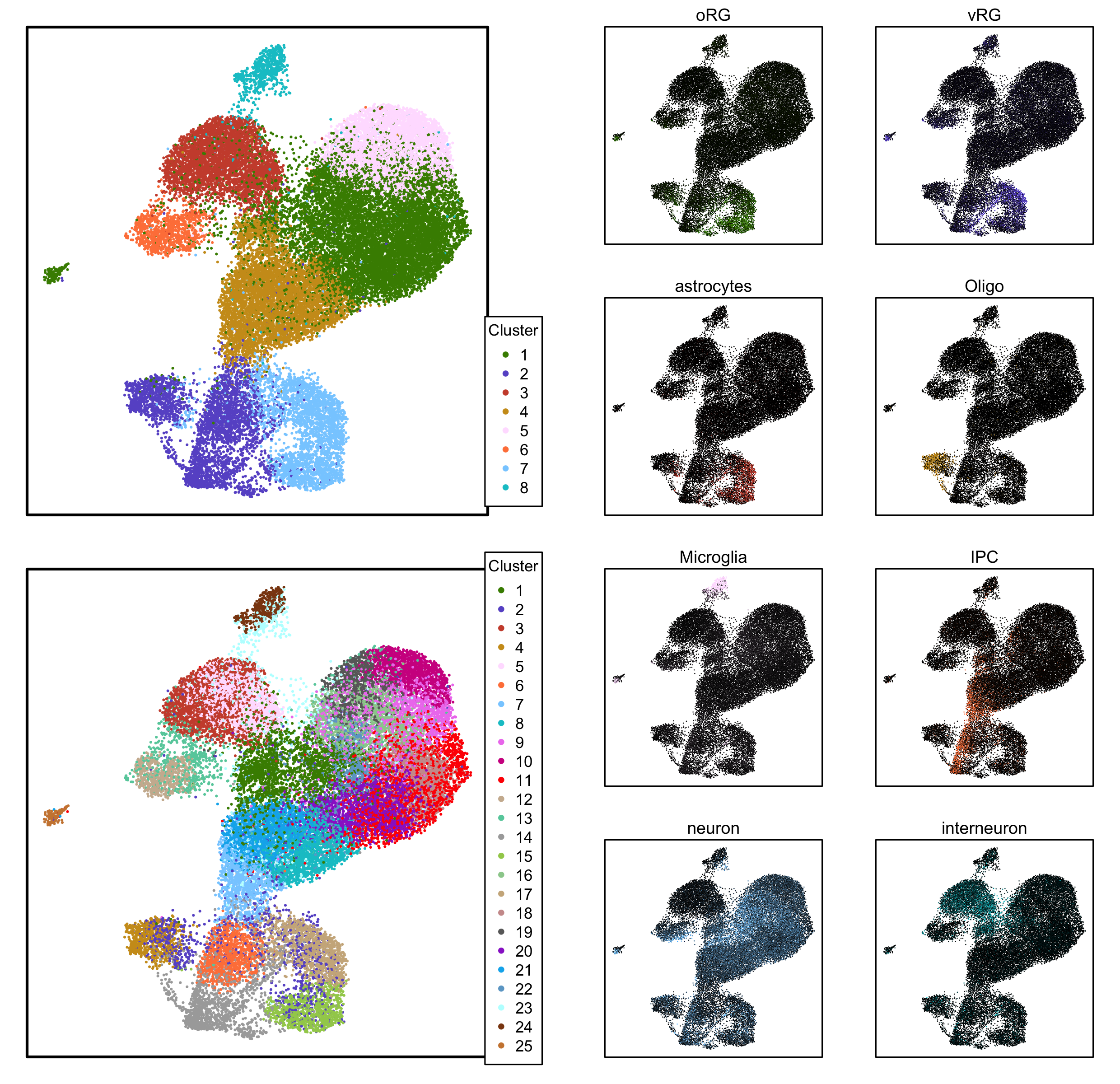}
\vspace{-1ex}
\caption{Projection, clustering, and validation, for the human cortical development data-set. {\normalfont UMAP projections from the Laplacian eigenspace (UMAP-LE) show GMM-LE clusters and mean expression levels of validating marker genes for cell-types of interest.}}\label{neuroDataFig}
\vspace{-2ex}
\end{figure}

The cell-type validation plots shown in Figure \ref{neuroDataFig} illustrate that the UMAP projection of the data-points in the Laplacian eigenspace (UMAP-LE) make biological sense. The neural stem-cell types oRG (outer radial glia) and vRG (ventricular radial glia) appear in separate yet close locations, and a trajectory can be discerned from these stem-cell types through IPC (intermediate progenitor cells) to neuron and interneurons. The GMM-LE clusters also closely correspond to the greatest colour intensity in the validation plots for specific cell-types. However, the number of clusters {\it K} is an important parameter to set in any clustering procedure. Because the validation for the clustering in the human cortical development data-set was based on marker genes for 8 different neural cell-types, we started with $K=8$ clusters for this data-set (Figure \ref{neuroDataFig}). However, this did not lead to the neural stem-cell types oRG (outer radial glia) and vRG (ventricular radial glia) being clustered separately, despite their apparent separation in the UMAP projection of the Laplacian eigenspace. Therefore, we tested $K\in\{5,10,15,20,25\}$, finding that $K\in\{20,25\}$ provided sufficient granularity to separate the neural stem-cell types oRG and vRG, with little difference in the clustering output between $K=20$ and $K=25$. However, at this level of granularity, it becomes clear that the definitions of `neuron' (i.e., excitatory neurons) and `interneuron' (i.e., inhibitory neurons) are too broad for our sets of validation markers, and that finer definitions of neuronal subtypes will be required to identify these (possibly novel) subtypes. We also compared these results from GMM-LE clustering against those obtained from the well-known Louvain clustering algorithm \cite{blondel2008fast} implemented in the industry-standard Seurat software package \cite{satija2015spatial}. The Seurat software with default settings chooses $K=22$, and therefore we used $K=22$ to compare results between the GMM-LE and Seurat-Louvain clustering methods. Table \ref{methodsCompTabNeuro} shows ratios of the mean log expression of the marker genes for one cell-type to the mean log expression for the markers for all the other cell-types. These are based on cell-type identity assigned to each cluster based on the set of markers with the greatest mean expression level in each cluster. For 6 out of 8 cell-types, GMM-LE clustering performs better; however as noted already, the worse performance for neurons and interneurons is likely to be due to the fact that these definitions are too broad for clustering at this level of granularity, and therefore these general cell-types span several clusters.

\begin{table}[h!]
\vspace*{-0.5ex}
\centering
{\small
\begin{tabular}{lrrrrrrrr}
   \hline
    Method & oRG & vRG & astrocytes & Oligo & Microglia & IPC & neuron & interneuron \\ 
   \hline
   GMM-LE & 1.43 & 1.16 & 1.29 & 1.56 & 1.44 & 1.56 & 1.17 & 1.41 \\
   Seurat-Louvain & 1.31 & 1.09 & 1.25 & 1.58 & 1.29 & 1.29 & 1.16 & 1.97 \\ 
   \hline
\end{tabular}
}
\vspace*{-1ex}
\caption{Comparison of GMM-LE clustering and Seurat-Louvain clustering in the human cortical development data-set. {\normalfont Ratios of the mean log expression of marker genes for the cell-type of interest to the other cell-types are shown.}}\label{methodsCompTabNeuro}
\vspace*{-1ex}
\end{table}

\subsection{Human embryonic development dataset: visualisation and clustering inferences}\label{embryoResSect}

\begin{figure}[!h]
\vspace{-1ex}
\centering
\hspace*{-4ex}
\raisebox{0.15\height}{\includegraphics[width=0.48\textwidth]{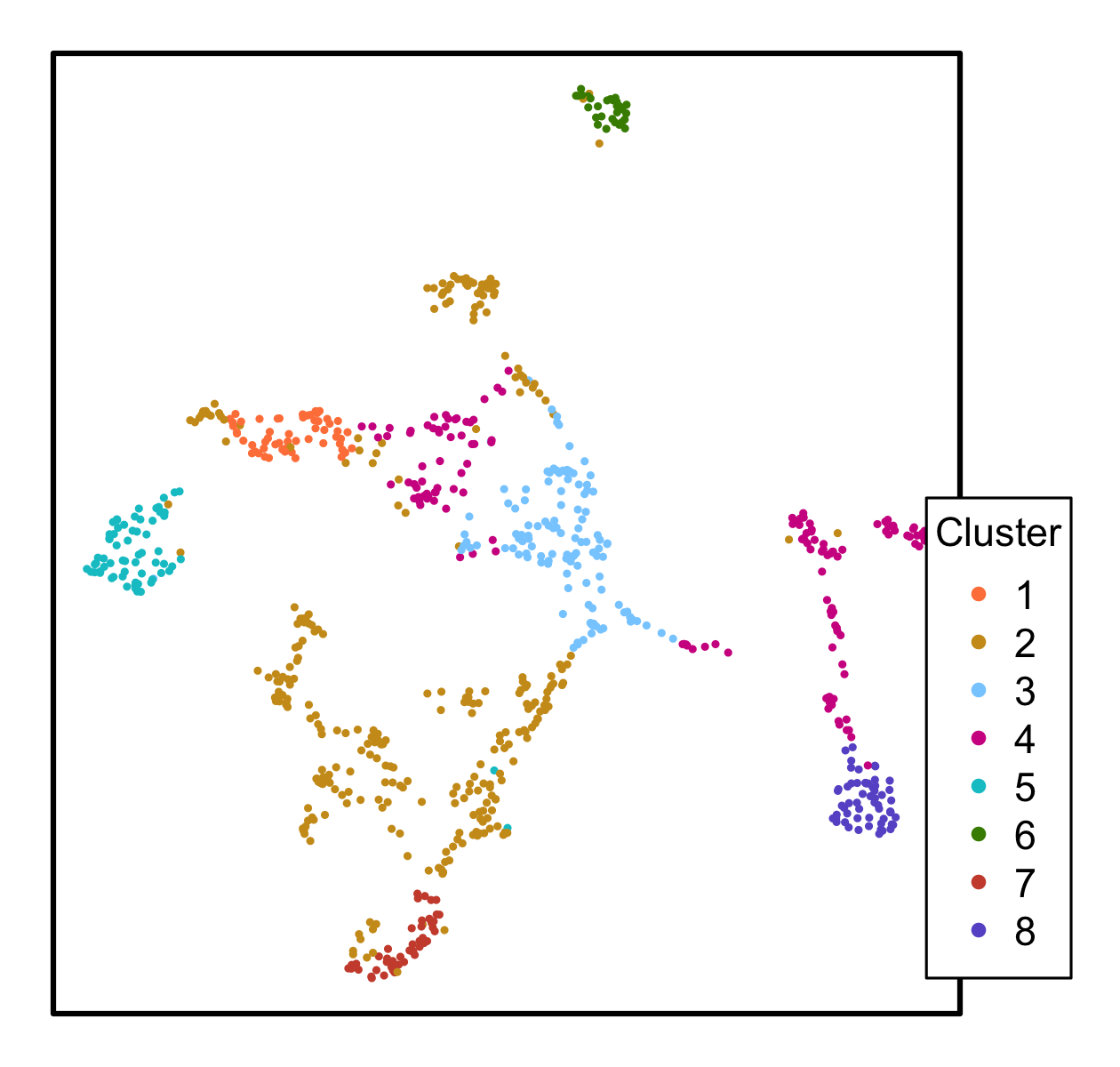}}
\hspace*{-2ex}
\includegraphics[width=0.54\textwidth]{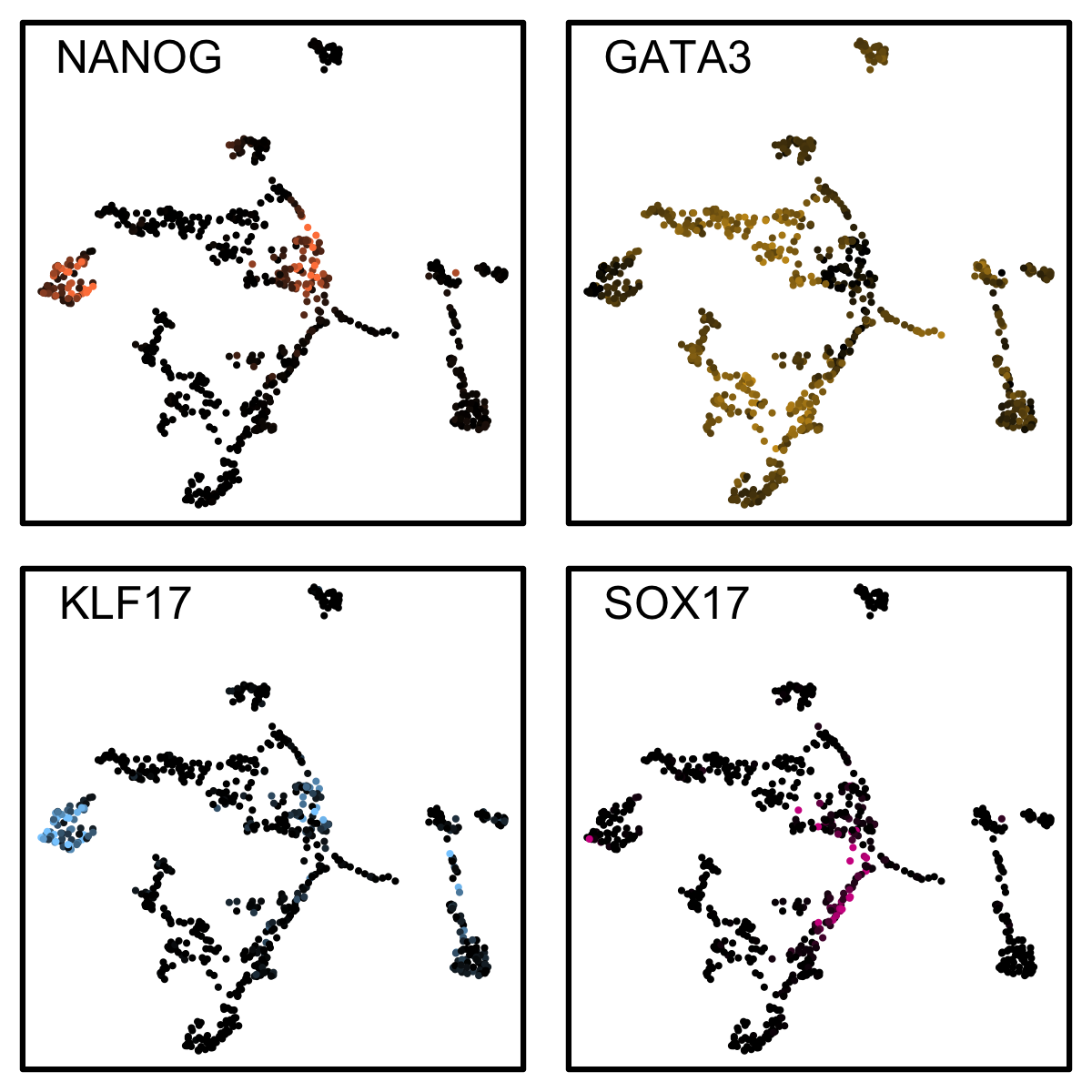}
\hspace*{-4ex}
\vspace{-1ex}
\caption{Projection, clustering, and validation, for the human embryonic development data-set. {\normalfont UMAP projections from the Laplacian eigenspace (UMAP-LE) show GMM-LE clusters and expression levels of validating marker genes for cell-types of interest.}}\label{embryoDataFig}
\vspace*{-2ex}
\end{figure}

Research on human embryos is highly restricted, due to ethical and regulatory considerations. This constrains sample sizes in generated data-sets, restricting inferences that can be obtained from these samples. If a novel clustering method is able to increase the number of cells obtained from an embryo that can be reliably identified as a specific cell-type of interest in the embryo such as epiblast cells, then it would be potentially very valuable to biological science researchers because it would increase the effective sample size for these precious cells that can be used for further research based on those data. Epiblast cells are precursors to the ectoderm layer, one of three primary germ layers of the embryo; epiblast cells go on to form neural and epithelial cells \cite{wainwright2010langman}. Using GMM-LE clustering, we are able to increase the number of cells that can be reliably identified as epiblasts from a well-known study of human embryos \cite{petropoulos2016single}, compared to the latest research in the field \cite{stirparo2018integrated}. To set the number of clusters {\it K} shown in Figure \ref{embryoDataFig}, we increased {\it K} until the epiblast cells formed a distinct cluster as shown by the UMAP projection of the Laplacian eigenspace (UMAP-LE), setting $K=8$. In Figure \ref{embryoDataFig}, cluster 5 identifies 68 epiblast cells according to expression of the marker genes NANOG and KLF17, and the absence of expression of GATA3 and SOX17 \cite{blakeley2015defining}. These 68 cells include 44 of 45 cells that were previously identified as epiblast, as well as very importantly 24 cells newly identified as epiblast cells by our proposed methodology. The full list of all 68 epiblast cells are given in \ref{epiCellIDs}.

\subsection{Breast cancer at-risk dataset: visualisation, clustering, and biomarker inferences}
The main cell-types of the epithelial lineage in breast tissue are luminal progenitor cells and mature luminal cells (thought to be the cells-of-origin of, respectively, hormone-receptor negative and positive breast cancers) as well as basal cells \cite{tharmapalan2019mammary}. As luminal progenitor cells are thought to be the cell-of-origin of highly aggressive triple negative breast cancers (TNBC), it is important to be able to identify these cells as accurately as possible, e.g., for quantifying their numbers in tissue samples from at-risk individuals. Important applications of accurately quantifying luminal progenitor cells include developing novel biomarkers and surrogate end-points for clinical trials of chemopreventative medicines \cite{bartlett2022antiprogestins}. 

\begin{figure}[!h]
\vspace{-1ex}
\centering
\hspace*{-4ex}
\raisebox{0.15\height}{\includegraphics[width=0.48\textwidth]{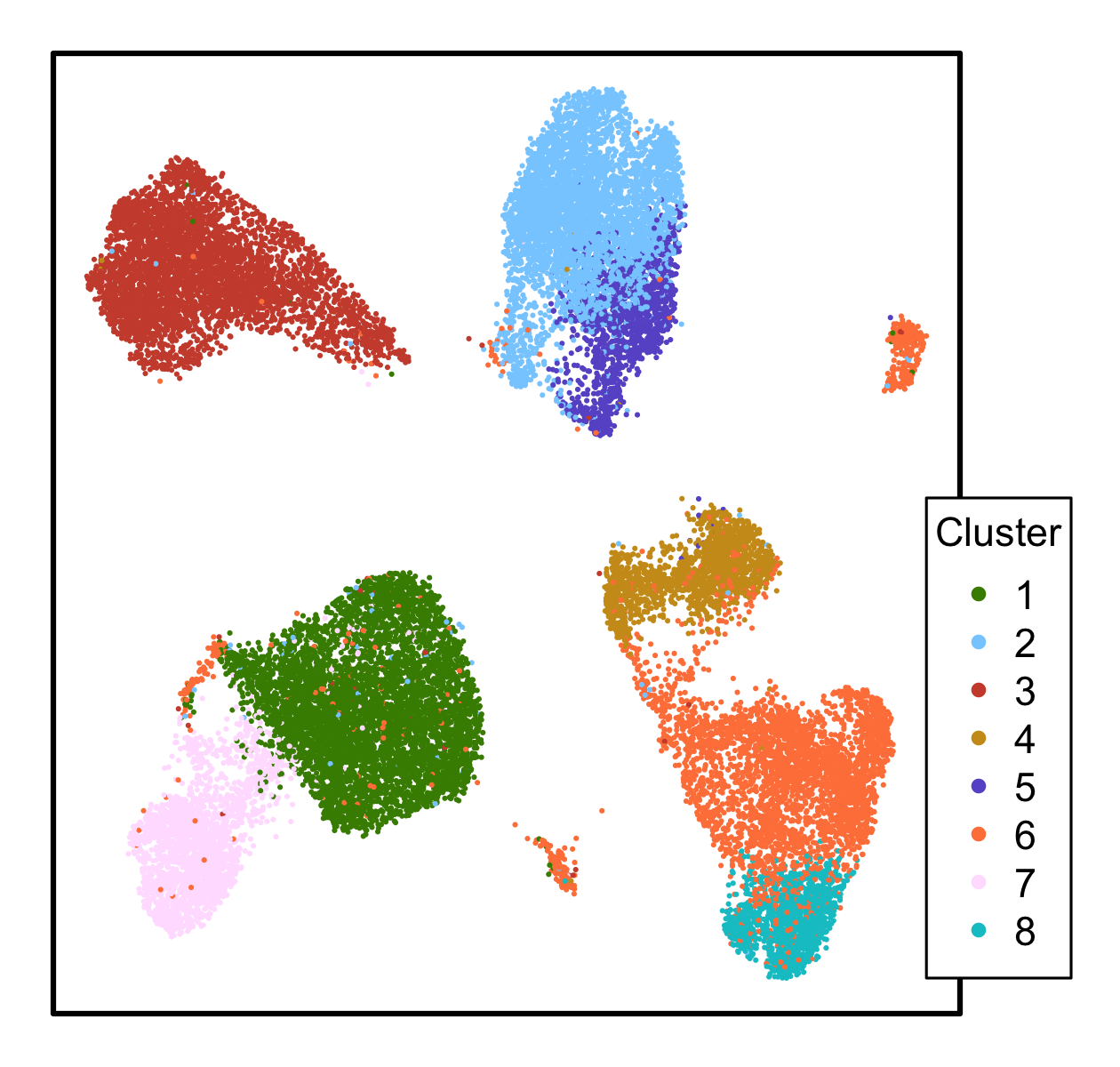}}
\hspace*{-2ex}
\includegraphics[width=0.54\textwidth]{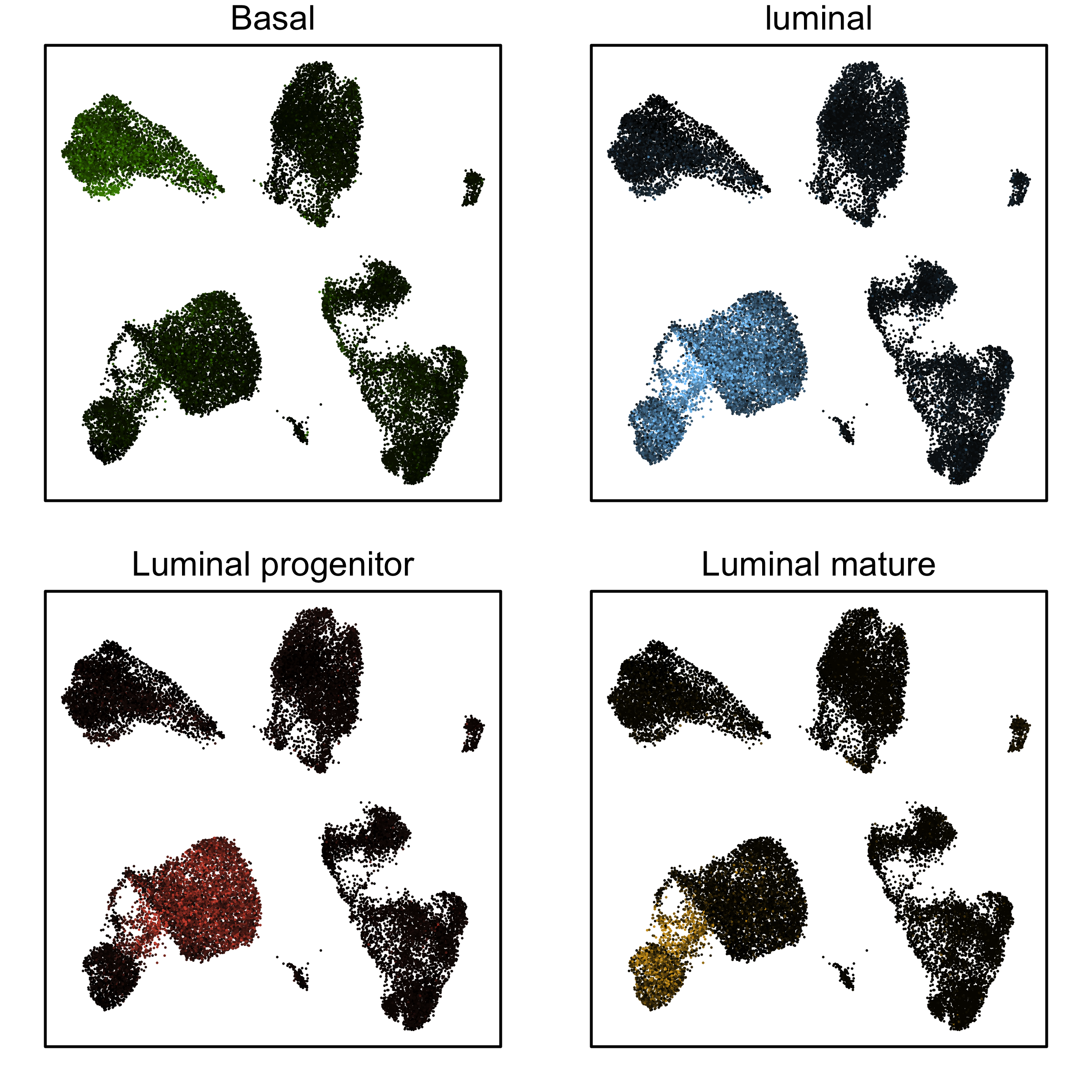}
\hspace*{-4ex}
\vspace{-1ex}
\caption{Projection, clustering, and validation, for the breast cancer at-risk data-set. {\normalfont UMAP projections from the Laplacian eigenspace (UMAP-LE) show GMM-LE clusters and mean expression levels of validating marker genes for cell-types of interest.}}\label{breastCancerDataFig}
\vspace{-2ex}
\end{figure}

We have previously shown that GMM-LE clustering improves identification of luminal progenitor cells in single-cell genomic data compared to Seurat-Louvain clustering \cite{bartlett2021inference}. In this study we identify the breast epithelial subtypes basal, luminal progenitor, and mature luminal cells in a different recently published data-set \cite{pal2021single}, again showing an improvement over Seurat-Louvain clustering (Table \ref{methodsCompTabBr}). As an additional comparison with Louvain clustering, we cluster these cells using the Louvain implementation available in the scikit-net package for Python \cite{bonald2020scikit}, again taking the genomic data-matrix as a bipartite network adjacency matrix. However, this approach finds 17 clusters, of which one is a giant cluster (comprising 97\% of the nodes/cells), indicating that the {\it k}-nearest-neighbour graph construction step in the Seurat-Louvain implementation is important for using Louvain clustering with genomic count data. This result remained unchanged with either the default `Dugue' or `Newman' settings in the scikit-network-Louvain implementation.

\begin{table}[h!]
\vspace*{-0.5ex}
\centering
{\small
\begin{tabular}{lrrr}
   \hline
    Method & Basal & Luminal progenitor & Luminal mature \\ 
   \hline
   GMM-LE & 3.86 & 5.54 & 10.4 \\
   Seurat-Louvain & 3.58 & 5.49 & 10.5 \\ 
   \hline
\end{tabular}
}
\vspace*{-1ex}
\caption{Comparison of GMM-LE clustering and Seurat-Louvain clustering in the breast cancer at-risk data-set. {\normalfont Ratios of the mean log expression of marker genes for the cell-type of interest to the other cell-types are shown.}}\label{methodsCompTabBr}
\vspace*{-1ex}
\end{table} 

The breast epithelial subtypes can be separated well in the UMAP projection of the data-points in the Laplacian eigenspace (UMAP-LE), and they can be separated well by GMM-LE clustering, shown in Figure \ref{breastCancerDataFig}. To carry out GMM-LE clustering here, we again increase the number of clusters {\it K} until the breast epithelial subtypes are well separated visually, choosing $K=8$. We also note that plotting the cells with positions according to their principal component scores for the top two dimensions (i.e., the first two rows of $\mathbf{V}$, as in \ref{markersPCAfig}) is less effective for visualising these cell types separately than generating a two-dimensional projection from the top 25 dimensions of $\mathbf{V}$.

\subsection{Breast cancer at-risk dataset: predicting long-term patient survival outcome}
\begin{figure}[!ht]
\centering
\vspace*{-1ex}
\hspace*{-2ex}
\includegraphics[width=1.05\textwidth]{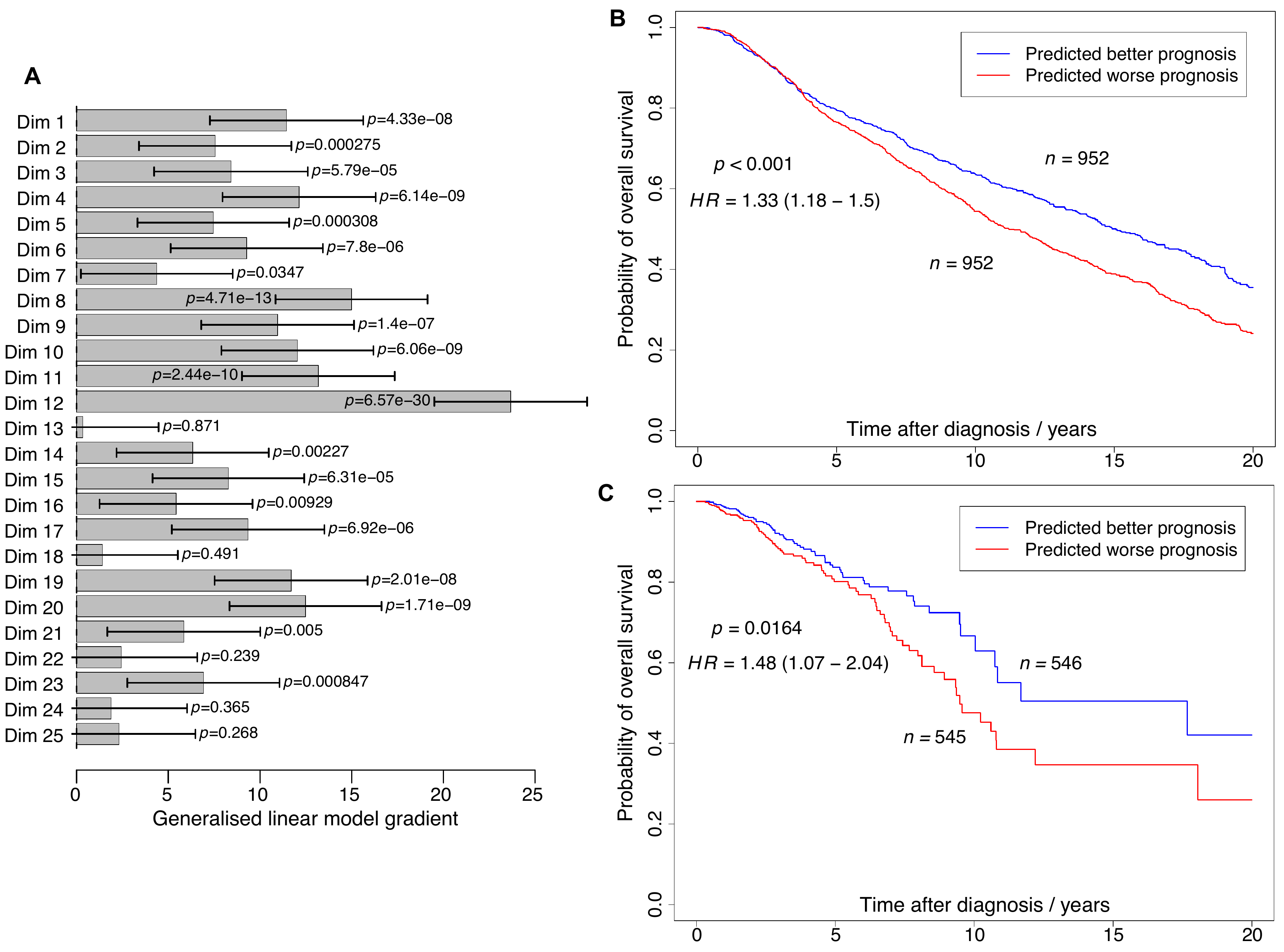}
\hspace*{-2ex}
\vspace{-3ex}
\caption{LE cancer biomarker inference, for the breast cancer at-risk data-set. {\normalfont(a) A generalised linear model (with logit link) was fitted to the top 25 RSVs (right singular vectors) as covariates (predictors), with {\it BRCA1} mutation status of the donor of the cell sample as the response; the 12th RSV of the LE is most strongly associated with {\it BRCA1} mutation status, as a surrogate measure of predisposition to aggressive cancer. (b) and (c) Kaplan-Meier curves show significant prediction of long-term patient survival outcome by a polygenic score (PGS) derived from the 12th LSV (left singular vector) of the LE, respectively in the independent METABRIC \cite{curtis2012genomic} and TCGA-BRCA \cite{cancer2012comprehensive} validation datasets.}}\label{breastCancerBiomarkerFig}
\vspace*{-2ex}
\end{figure}

As a further application of our proposed methodology, we develop a novel way to discover biomarkers of risk of aggressive cancer. We do so by testing the association of each dimension of the LE with a known genetic marker of predisposition to the most aggressive breast cancers, a mutation in the {\it BRCA1} ({\it FANCS}) gene \cite{tharmapalan2019mammary}. Individuals with a mutation in the {\it BRCA1} ({\it FANCS}) gene have a much increased risk of breast cancer during their lifetime \cite{kuchenbaecker2017evaluation}. This means that cells from the healthy tissue of {\it BRCA1} mutation carriers may have some characteristics that predispose some of these cells to ultimately transform into aggressive cancer \cite{bartlett2022antiprogestins,williams2024luminal}. These characteristics of cancer predisposition would also be expected to be reflected in the gene expression profiles of these cells. Therefore, we compare scores derived from each dimension of the LE transformed latent positions in healthy cells from {\it BRCA1} mutation carriers (BRCAmut) with those from cells from {\it BRCA1} wild-type (WT). To do so, we fit a generalised linear model with logit link-function to the top 25 RSVs $\mathbf{V}\in\mathbb{R}^{n\times 25}$ as covariates (predictors), with {\it BRCA1} mutation status of the donor of the cell sample as the response. The absolute value of the gradient of the fitted linear model for each of these dimension is shown in Figure \ref{breastCancerBiomarkerFig}a (with {\it z}-test {\it p}-values); dimension 12 has the largest absolute gradient and smallest {\it p}-value, and hence is selected for the LE biomarker. 

To identify the biomarker from LE dimension 12, we first check the signs of the relevant elements of $\mathbf{U}$ and $\mathbf{V}$, to separate assortative from disassortative effects (as discussed in Section \ref{modSec}). \ref{pcScoresFig} illustrates that if dimension $l\in\{1,...,d\}$ identifies certain characteristics of cells $j\in\{1,...,n\}$, then this corresponds to negative RSV values $V_{j,l}$ for these cells. For example, dimensions 1 and 3 identify luminal and basal cells (respectively) by comparison with Figure \ref{breastCancerDataFig}, and these cells are coloured strongly in red (indicating negative values) for these dimensions in \ref{pcScoresFig}. Dimension 12 is most strongly associated with {\it BRCA1} mutation status and thus predisposition to aggressive cancer, and also shows a concentration of cells marked in red (indicating negative values) corresponding to cells from {\it BRCA1} mutation carriers by comparison with \ref{qcBrFig}. Furthermore, the linear model coefficient (in the glm described above) is negative for the 12th RSV. Therefore, we choose the most negative elements of the corresponding LSV as those that would identify the genes that contribute most to the assortative effect for LE dimension 12. This is because data value $X_{i,j}$ for gene $i$ in cell $j$ with corresponding Laplacian element $\mathcal{L}_{i,j}$ will have a large contribution from dimension 12 in the SVD $\boldsymbol{\mathcal{L}}=\mathbf{U}\mathbf{S}\mathbf{V}^\top$ when the signs are the same for the $\mathbf{U}$ and $\mathbf{V}$ elements $U_{i,12}$ and $V_{j,12}$. We therefore select the 100 genes corresponding to the most negative elements of the 12th LSV as a biomarker of risk of aggressive cancer, to calculate a polygenic expression score (PGS). We validate this PGS using two independent previously-published data-sets of gene expression profiles from respectively 1904 tumour samples (the `METABRIC' validation dataset \cite{curtis2012genomic}) and 1091 tumour samples (the `TCGA-BRCA' validation dataset \cite{cancer2012comprehensive}). The Kaplan-Meier curves shown in Figure \ref{breastCancerBiomarkerFig}b-c show that this LE cancer biomarker significantly predicts long-term patient survival ($p<0.001$ and $p=0.0164$ respectively, Mantel-Haenszel test), comparing good and bad prognostic groups divided by the median PGS, in both these independent validation datasets. 

\section{Conclusions}
In this paper we have presented a novel first-principles method of modelling single-cell genomic count data, by proposing to model the observed numbers of mRNA molecules in specific cells with a bipartite multi-edge network. This view of single-cell genomic data opens up a wealth of new modelling approaches for this type of data based on stochastic networks theory. Our proposed methodology is also closely related to well-established and highly successful methods for clustering these data based on community detection, such as Louvain clustering \cite{blondel2008fast}. The work we have presented here shows the promise of studying the eigenspace of the graph-Laplacian (the `Laplacian eigenspace', LE), to identify and represent known and novel cell-types. Based on this, we have proposed a new method of visualising single-cell genomic data based on the UMAP projection of the representation of the data-points in the Laplacian eigenspace (UMAP-LE), and we have proposed GMM-LE clustering, or `Gaussian mixture modelling in the Laplacian eigenspace', for genomic data. We show that UMAP-LE visualisations correspond to existing biological knowledge in several data-sets. We have also identified a component of the LE that is associated with predisposition towards aggressive breast cancer, and have validated a biomarker extracted from this LE component, showing significant association with long-term patient survival outcome. 

We have validated our proposed methodology with real world data-sets that are relevant to timely problems in several biomedical science domains. We note that an alternative strategy for validating our proposed methodology would have been to use artificially constructed data-sets consisting of mixtures of pre-determined cell-types with known ground-truth. However, we believe that it is more realistic, and more useful to wet lab scientists who may wish to use our proposed methodology, to demonstrate its effectiveness in the context of real-world datasets obtained directly from single cells that have been dissociated together from the same live tissue. It is unknown what changes may occur to the transcriptomic profiles of cells, if cells of different types are mixed together in an unnatural way in order to construct a reference data-set. Hence such a procedure may introduce additional unwanted sources of variation in the data, potentially leading to confounding. Instead, by using carefully chosen sets of marker genes provided by domain experts, we have been able to provide a compelling validation of our proposed methodology in more realistic settings. Furthermore, this setup also allows novel cell-types to be discovered.

There are several outstanding questions still to address in this research. In particular, choosing the number of clusters for a particular data-set is a challenging problem, which we have addressed heuristically for each data-set here. Developing an automatic method for choosing the number of clusters for GMM-LE clustering is a priority. A straightforward approach to this problem is to set the number of clusters as one greater than the dimensionality of the Laplacian eigenspace that is used \cite{riolo2014first}. We note that the latent dimensionality of the eigenspace of the graph-Laplacian can be estimated manually by identifying an eigengap in the ordered plot of singular values, or by using existing theory based on scree plots \cite{zhu2006automatic}. Recent findings related to stochastic network theory provide refined methodology to choose the optimum number of clusters to estimate in the Laplacian eigenspace via the Bayesian Information Criterion (BIC) \cite{priebe2019two}. Building on existing theory for bootstrapping network data with latent structure \cite{levin2019bootstrapping} will enable estimates of uncertainty to be calculated for estimators of the number of latent clusters, as well as quantifying uncertainty on the estimated Laplacian eigenspace. Finally, we note that we have previously proposed a novel method for estimating a latent space in which different cell-types or clusters can be separated well, based on the Mahalanobis distance \cite{bartlett2021inference}. It will be interesting to investigate how that idea relates to the estimate of the Laplacian eigenspace, and how it may be refined for identifying different cell-types from single-cell genomic data.

In summary, the methodological advances that we present here provide mathematical technologies that have the potential for large impact in biomedical science. This is achieved by providing theoretical justification for and improvement on existing and widely adopted methodology for identifying important cell-types in single-cell genomic data.

\section*{Data sources}
\vspace{-1ex}
The human cortical development data-set is available from the NeMO (Neuroscience Multi-Omic) data archive from URL:\\
\noindent\url{https://data.nemoarchive.org/biccn/grant/u01_devhu/kriegstein/transcriptome/scell/10x_v2/human/processed/counts/}\\
\noindent
Gene expression data for 33694 RNA transcripts in 44885 cells were downloaded, corresponding to cells from the V1 area of the visual cortex from four embryos at 20-22 gestation weeks. After quality control and filtering, data for 15735 features (RNA transcripts) in 31073 samples (cells) remained, and were carried forward for the subsequent modelling and analysis. 

The human embryonic development data-set is available from the ArrayExpress repository under accession number E-MTAB-3929, from URL:\\
\noindent\url{https://www.ebi.ac.uk/biostudies/arrayexpress/studies/E-MTAB-3929}\\
\noindent
Gene expression data for 26178 RNA transcripts in 1529 cells were downloaded, corresponding to cells from the blastocyst stage of embryonic development.  After quality control and filtering to retain cells from embryos at 5-7 days post-fertilisation, data for 20407 features (RNA transcripts) in 1258 samples (cells) remained, and were carried forward for the subsequent modelling and analysis. 

The breast cancer at-risk data-set is available from the Gene Expression Omnibus (GEO) repository under accession number GSE161529, from URL:\\
\noindent\url{https://www.ncbi.nlm.nih.gov/geo/query/acc.cgi?acc=GSE161529}\\
\noindent
Gene expression data for 33538 RNA transcripts in 26272 cells were downloaded, corresponding to cells from the epithelial lineage and supporting cells. After quality control, data for 14835 features (RNA transcripts) in 17730 samples (cells) remained, and were carried forward for the subsequent modelling and analysis.

The METABRIC breast cancer validation dataset is available from the European Genome-Phenome Archive (EGA) under accession number EGAS00000000083, from URL:\\ 
\noindent\url{https://ega-archive.org/studies/EGAS00000000083}\\
\noindent
Gene expression data for 24368 genes and 1904 patient volunteers were downloaded, together with patient survival outcome data for all volunteers in the study. There were 10 missing values in the genomic data-set, which were replaced with {\it K}-NN imputation, with $K=5$.

The TCGA-BRCA breast cancer validation dataset is available from the NIH GDC Data Portal under project TCGA, from URL: \url{https://portal.gdc.cancer.gov}\\\noindent
Gene expression data for 20499 transcripts and 1091 patient volunteers were downloaded, together with patient survival outcome data for all volunteers in the study. There were no missing values in the data-set.

\section*{References} 
\printbibliography[heading=none]

\section*{Acknowledgements}
\vspace{-1ex}
The work carried out by TE Bartlett was partly supported by MRC grant (MR/P014070/1). The authors acknowledge the use of the UCL High Performance Computing Facilities, and associated support services, in the completion of this work. The results published here are in part based on data generated by the TCGA Research Network: \url{https://www.cancer.gov/tcga}

\section*{Competing interests}
\vspace{-1ex}
The authors declare that there are no competing interests.

\section*{License}
\vspace{-1ex}
This manuscript is available under a Creative Commons CC BY-NC License, as described at\\ \url{https://creativecommons.org/licenses/by-nc/4.0/}

\clearpage

\setcounter{page}{1}

\setcounter{figure}{0}
\setcounter{table}{0}
\renewcommand{\thefigure}{Figure S\arabic{figure}}
\renewcommand{\figurename}{}
\renewcommand{\thetable}{Table S\arabic{table}}
\renewcommand{\tablename}{}

\vspace{-2ex}
\section*{Supplementary Information A: Data pre-processing}
For quality control, high quality features (RNA transcripts) and samples (cells) were retained by filtering out any RNA transcripts with non-zero counts in fewer than 50 cells, and by filtering out any cells with non-zero counts for fewer than 750 RNA transcripts, filtering out cells for which the most highly expressed transcript excluding MALAT1 account for at least 10\% of the total counts, filtering out cells for which at least 10\% of the total counts correspond to transcripts encoding mitochondrial proteins, and filtering out cells for which at least 50\% of the total counts correspond to transcripts encoding ribosomal proteins. For the human embryonic data-set, features were instead filtered out if they have non-zero counts in fewer than 10 samples (cells), noting the smaller sample size for this data-set; also for he human embryonic data-set no filtering was carried out based on transcripts encoding mitochondrial proteins because they did not appear in the data-set. For the human cortical development data-set and the breast cancer at-risk data-set, batch correction using the COMBAT software \cite{johnson2007adjusting} was also applied to remove inter-individual differences. 

After quality control, the top features were selected for dimension reduction, visualisation, and clustering, according to an estimator of the amount of biological variability compared to technical variability \cite{mccarthy2012differential} in each feature $i\in\{1,...,p\}$ defined as $\mathrm{log}V_i/\mathrm{log}m_i$, where $V_i$ and $m_i$ are the empirical variance and mean for feature {\it i}, respectively. The top 2000 features were retained in this way for the human cortical development data-set and the breast cancer at-risk data-set, and the top 5000 features were retained for the embryonic development data-set. We note that this procedure in practice approximates a standard technique for obtaining the features with the greatest biological variability in which a line of best fit is found for a plot of $\mathrm{log}V_i$ against $\mathrm{log}m_i$, and the features that are furthest above this line are then selected. This practice is based on the assumption that the data for feature {\it i} will follow a Poisson distribution when only technical variation is present meaning that $V_i\approx m_i$, and will be overdispersed when biological variation is present meaning that $V_i\approx m_i+\phi_i m_i^2$, where $\phi_i>0$ is an overdispersion parameter for feature {\it i} \cite{mccarthy2012differential}. The Laplacian eigenspace is then estimated as the space defined by the eigendecomposition of the graph-Laplacian, according to the eigenvectors that correspond to the eigenvalues that explain at least 1\% of the total variance of the graph-Laplacian representation of the data-points.

\vspace{-1ex}
\section*{Supplementary Information B: Supplementary Tables}
\vspace{-2ex}
\begin{table}[!ht]
\begin{minipage}{0.35\linewidth}
\centering
\small
\vspace{-8ex}
\begin{tabular}{|c|r|r|}
    \hline 
    & With FS & Raw data \\
        \hline 
HCD & 0.9998 & 0.8869 \\
        \hline 
HED & 0.9992 & 0.7340 \\ 
        \hline 
BCAR & 1.0000 & 0.8835 \\
    \hline 
 \end{tabular}
\caption{Correlation coefficients comparing library size with the top RSV of the SVD of the graph-Laplacian. {\normalfont The top RSV corresponds to the largest singular value. Spearman correlation was used; all reported values correspond to $p<2.2\times10^{-16}$ (asymptotic {\it t}-test approximation). Abbreviations: Feature selection (FS); Human cortical development (HCD); Human embryonic development (HED); Breast-cancer at-risk (BCAR).}} \label{topRSVcor}
\end{minipage}
\hfill
\begin{minipage}{0.6\linewidth}
\centering
\small
\begin{tabular}{|c|c|c|c|}
    \hline
    E6.10.1046 & E6.17.1588 & E6.9.1024 & E7.17.1348 \\ 
    \hline
    E6.10.1048 & E6.17.1611 & E7.10.760 & E7.17.1353 \\ 
    \hline
    E6.10.1049 & E6.17.1612 & E7.10.761 & E7.6.262 \\ 
    \hline
    E6.10.1050 & E6.17.1617 & E7.10.762 & E7.6.264 \\ 
    \hline
    E6.10.1051 & E6.22.1852 & E7.10.764 & E7.8.311 \\ 
    \hline
    E6.10.1052 & E6.22.1866 & E7.10.766 & E7.8.312 \\ 
    \hline
    E6.10.1055 & E6.8.791 & E7.10.768 & E7.8.317 \\ 
    \hline
    E6.10.1056 & E6.8.792 & E7.12.858 & E7.8.318 \\
    \hline 
    E6.10.1057 & E6.8.794 & E7.12.861 & E7.8.329 \\ 
    \hline
    E6.10.1058 & E6.8.801 & E7.12.869 & E7.8.333 \\ 
    \hline
    E6.12.1274 & E6.8.802 & E7.12.871 & E7.8.343 \\ 
    \hline
    E6.12.1289 & E6.8.813 & E7.14.895 & E7.9.547 \\
    \hline 
    E6.13.1378 & E6.8.815 & E7.17.1334 & E7.9.554 \\
    \hline 
    E6.13.1381 & E6.8.816 & E7.17.1335 & E7.9.556 \\ 
    \hline
    E6.13.1389 & E6.8.820 & E7.17.1342 & E7.9.562 \\
    \hline 
    E6.16.1501 & E6.8.821 & E7.17.1346 & E7.9.569 \\ 
    \hline
    E6.17.1586 & E6.9.1022 & E7.17.1347 & E7.9.571 \\ 
    \hline
\end{tabular}
\caption{The 68 human embryo epiblast cells identified in Section \ref{embryoResSect}. {\normalfont The listed cell IDs correspond to the IDs given in the original study that generated these data \cite{petropoulos2016single}.}} \label{epiCellIDs}
\end{minipage}
\end{table}

\vspace{-2ex}
\section*{Supplementary Information C: Supplementary Figures}

\begin{figure}[!h]
\vspace{-3ex}
\centering
\includegraphics[width=0.6\textwidth]{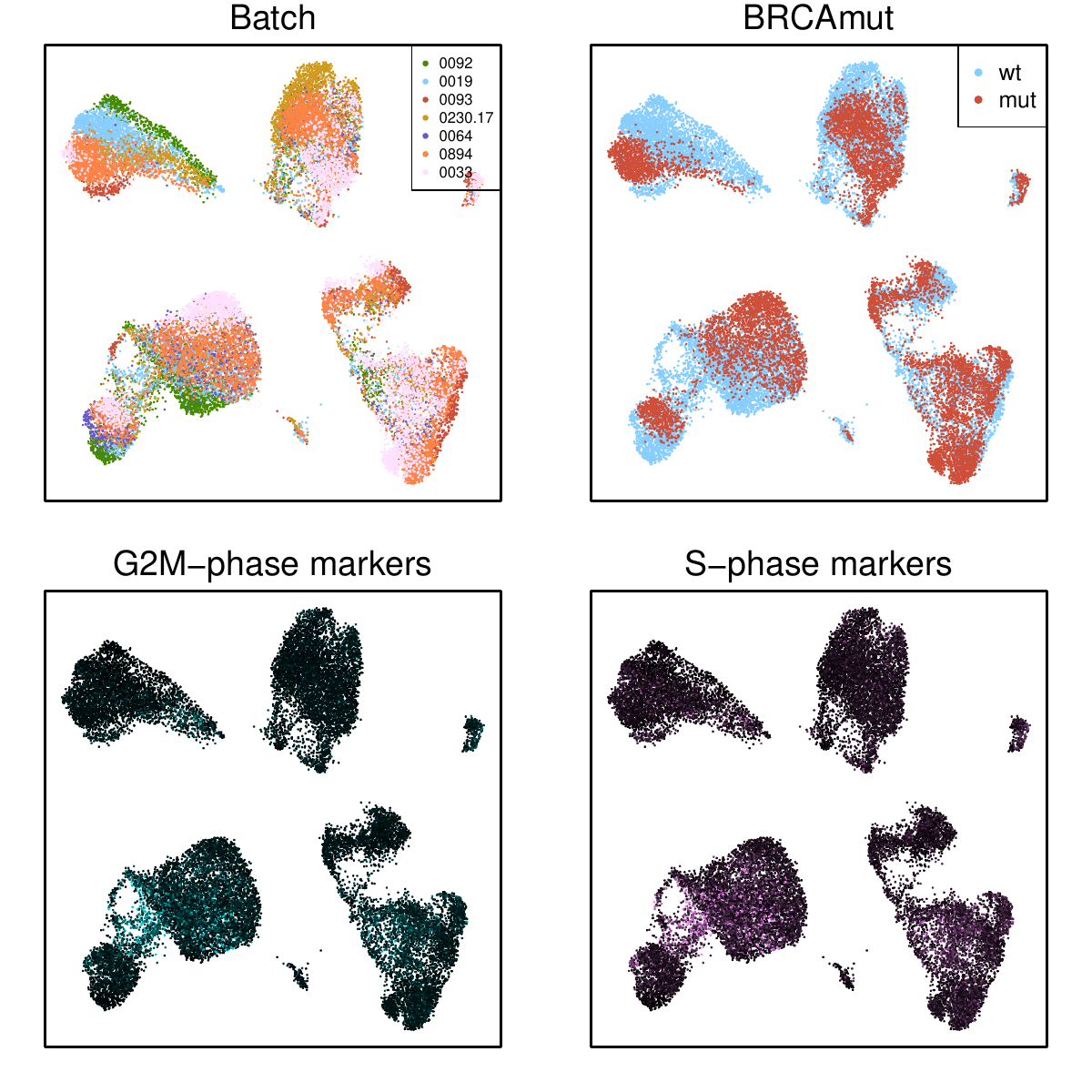}
\vspace{-1ex}
\caption{Quality control, batch, and demographic info, for the breast-cancer at-risk data-set. {\normalfont UMAP projections from the Laplacian eigenspace (UMAP-LE) display cells $j\in\{1,...,n\}$ coloured according to batch, {\it BRCA1} mutation status, or mean expression level of cell-cycle marker genes.}}\label{qcBrFig}
\vspace{-1ex}
\end{figure}

\begin{figure}[!h]
\vspace{-1ex}
\centering
\includegraphics[width=0.6\textwidth]{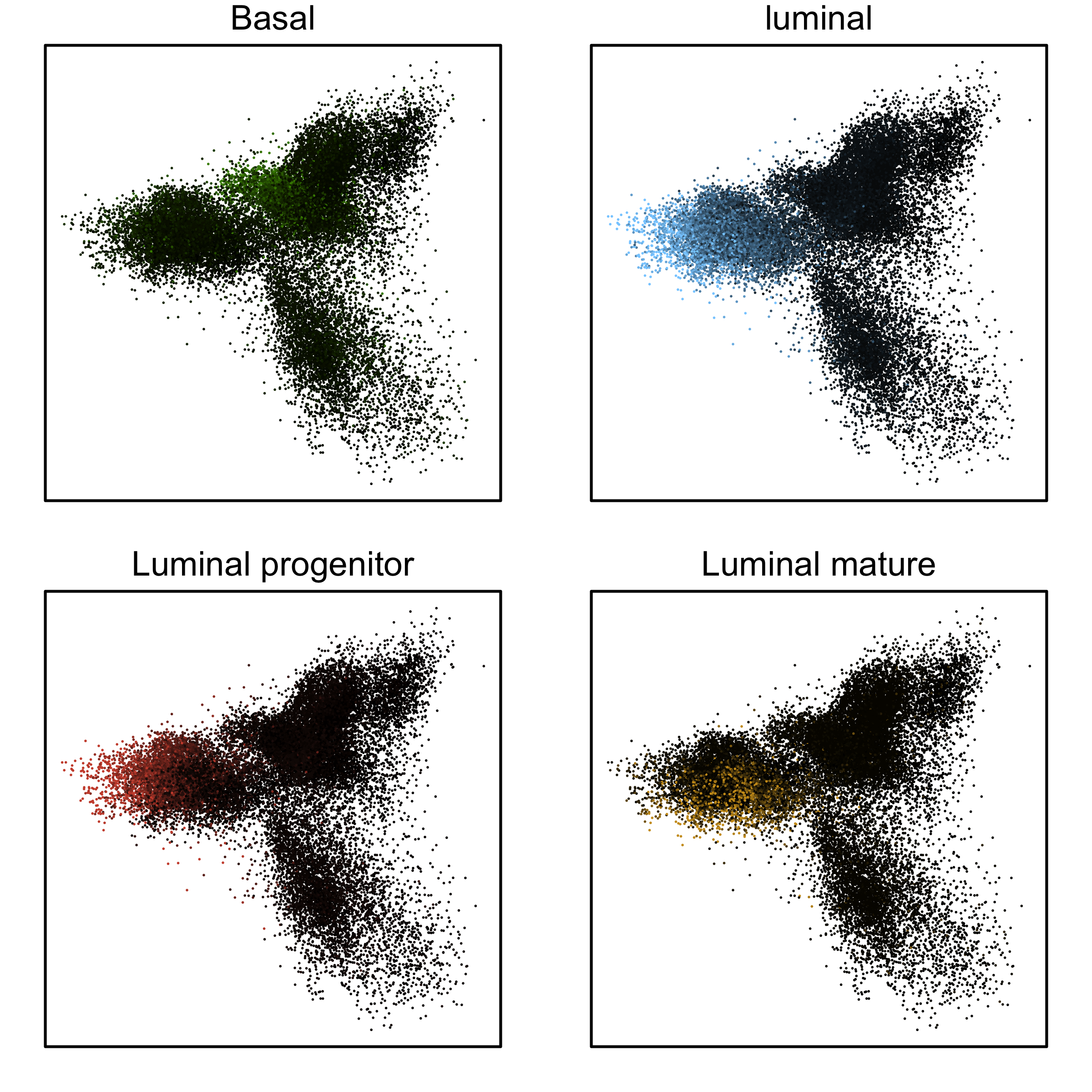}
\vspace{-1ex}
\caption{The top two Laplacian eigenspace dimensions as two-dimensional transformed latent positions, for the breast-cancer at-risk data-set. {\normalfont The top two rows of $\mathbf{V}$ display cells $j\in\{1,...,n\}$ coloured according to mean expression levels of validating marker genes for cell-types of interest..}}\label{markersPCAfig}
\vspace{-7ex}
\end{figure}

\begin{figure}[!h]
\vspace{-1ex}
\centering
\hspace*{-3ex}
\includegraphics[width=1.05\textwidth]{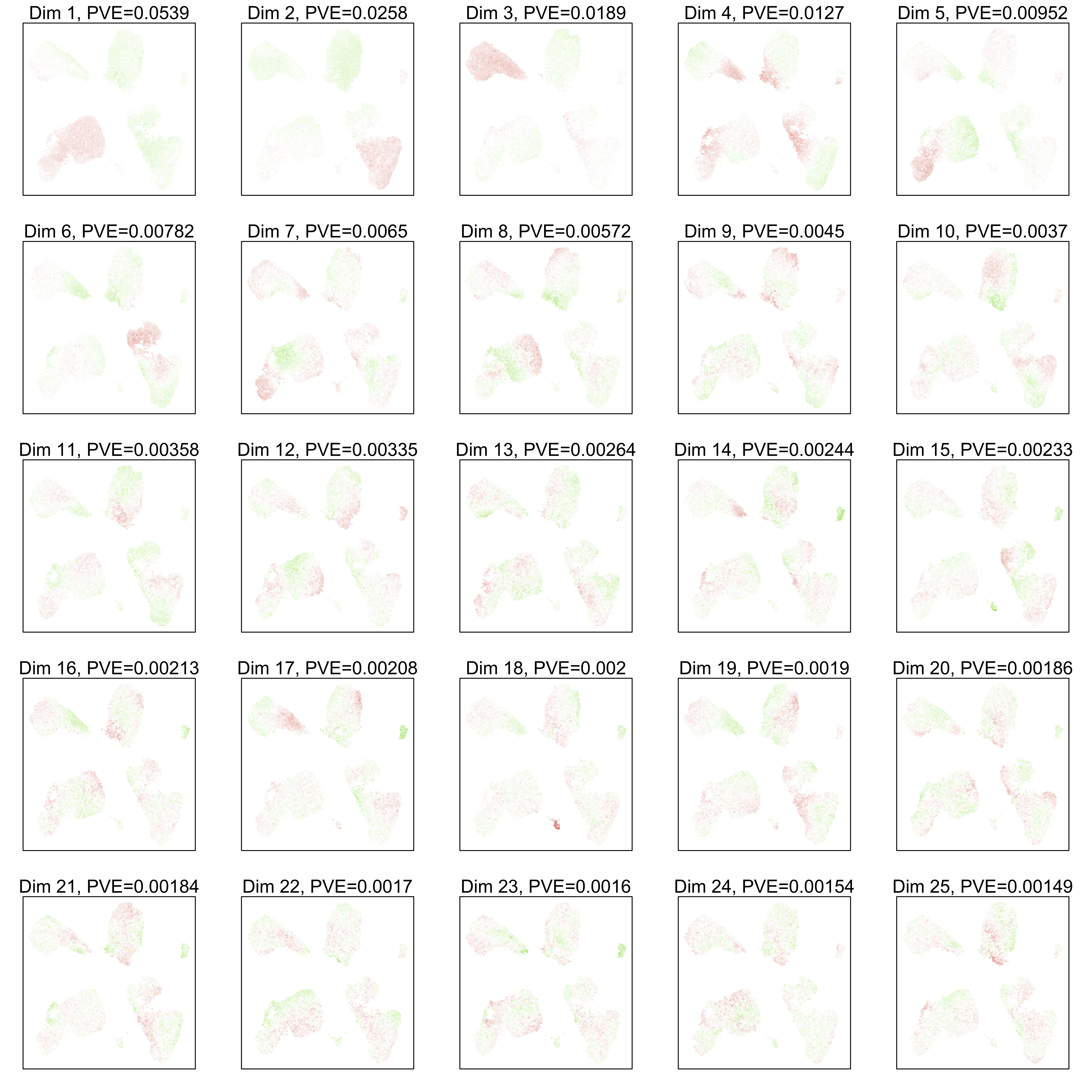}
\vspace{-1ex}
\caption{Laplacian eigenspace / transformed latent position dimensions as PC (principal component) scores, for the breast-cancer at-risk data-set. {\normalfont UMAP projections from the Laplacian eigenspace (UMAP-LE) display cells $j\in\{1,...,n\}$ coloured according to their RSV value $V_{j,l}$, for each dimension $l\in\{1,...,d\}$. Green and red indicate positive and negative values of $V_{j,l}$ (respectively), with colour intensity proportional to magnitude.}}\label{pcScoresFig}
\vspace{-1ex}
\end{figure}

\end{document}